\documentclass[aps,pre,preprint,floatfix,superscriptaddress]{revtex4}
\usepackage[utf8]{inputenc}
\usepackage{amsmath}
\usepackage{graphicx}  
\usepackage{amssymb}
\usepackage{epsfig}
\usepackage{natbib}
\usepackage{grffile}
  
\newcommand{\BE}{\begin{equation}}
\newcommand{\EE}{\end{equation}}

\usepackage[caption=false]{subfig}
\usepackage{graphicx}
\usepackage{dcolumn}
\usepackage{bm}

\def\bq{\begin{equation}}
\def\eq{\end{equation}}

\begin{document}


\title{A statistical study of gyro-averaging effects in a reduced model of drift-wave transport}
\author{J. D. da Fonseca}
\email{jfonseca@if.usp.br}
\affiliation{Physics Institute, University of São Paulo \\ São Paulo, SP, 5315-970, Brazil}

\author{D. del-Castillo-Negrete}
\email{delcastillod@ornl.gov}
\affiliation{Oak Ridge National Laboratory \\ Oak Ridge, TN, 37831-8071, USA}

\author{I. M. Sokolov}
\email{sokolov@physik.hu-berlin.de}
\affiliation{Physics Institute, Humboldt University \\ Berlin, Germany}

\author{I. L. Caldas}
\email{ibere@if.usp.br}
\affiliation{Physics Institute, University of São Paulo \\ São Paulo, SP, 5315-970, Brazil}

\date{\today}

\begin{abstract}
A statistical study of finite Larmor radius (FLR) effects on transport driven by electrostatic drift-waves is presented. 
The study is based on a reduced discrete Hamiltonian  dynamical system known as the 
gyro-averaged standard map (GSM). In this system,  FLR effects are incorporated through the gyro-averaging of 
a simplified weak-turbulence model of electrostatic fluctuations. Formally, the GSM is a modified version of the  standard map in which the perturbation amplitude, $K_0$, becomes $K_0 J_0(\hat{\rho})$, where $J_0$ is the zeroth-order Bessel function and $\hat{\rho}$ is the Larmor radius. Assuming a Maxwellian probability density function (pdf) for $\hat{\rho}$, we compute analytically and numerically the pdf and the cumulative distribution function of the effective drift-wave perturbation amplitude $K_0 J_0(\hat{\rho})$.   
Using these results we compute the probability of loss of confinement (i.e., global chaos), $P_{c}$, and the probability
of trapping in the main drift-wave resonance, $P_{t}$. It is shown that $P_{c}$ provides an upper bound for the 
escape rate, and that $P_{t}$ provides a good estimate of  the particle trapping rate. The analytical results are compared with direct numerical Monte-Carlo simulations of particle transport. 
\end{abstract}



\maketitle

\section{Introduction}

Particle transport  in magnetically confined plasmas devices such as tokamaks and stellarators is
commonly studied using the $\mathbf{E}\times\mathbf{B}$ drift approximation of the 
guiding center motion  that neglects  
finite Larmor radius (FLR) effects \cite{white}. 
Although in some cases this is a valuable approximation, it might break down when studying transport of 
$\alpha$-particles in burning plasmas or transport of fast particles in the  presence of  magnetic and/or electric field variations on the scale of the Larmor radius. 

A problem of particular interest is when the transport process is driven by electrostatic drift-waves \cite{Horton99}. In the simplest version of this problem,  the magnetic field is assumed constant. However, despite this  simplification, this problem is still quite challenging because, in principle, the electrostatic potential needs to be obtained from the self-consistent plasma dynamics, e.g., from the solution of the Hasegawa-Mima or the Hasegawa-Wakatani equations. 

One strategy to advance our understanding of this problem without the need of invoking the solution of turbulence models is to use a simplified description
for the electrostatic potential. This approach, which is the one adopted in the present paper, has opened the possibility of studying 
$\mathbf{E}\times\mathbf{B}$ transport using advanced methods and ideas from dynamical systems (see for example 
Refs.~\cite{kleva84, Horton85, Pettini,  Del-castillo-negrete00, Markus08} and reference therein). 
In particular, based on a weak-turbulence type assumption, we model the drift-wave electrostatic potential as a superposition of modes that allows to reduce the problem to a discrete Hamiltonian map. 
References following this approach include \cite{Horton98} where  the study of drift-wave transport was reduced to the study of a $2$-dimensional area preserving map. However, going beyond  these works, here we include FLR effects in the discrete Hamiltonian map description.

FLR effects on $\mathbf{E}\times\mathbf{B}$ transport have also been studied in Refs.~\cite{Annibaldi02,Manfredi96,Manfredi97} where the authors investigated
particle transport in numerical simulations of electrostatic turbulence and showed that FLR effect inhibit 
diffusive transport.
The role of FLR effects on non-diffusive chaotic transport, and fractional diffusion in particular, was addressed in Ref.~\cite{gustafson}. More recently, Refs.~\cite{del-castillo-martinell2012,del-castillo-martinell2013} studied the Larmor radius dependence of the phase space topology and the gyro-averaged induced suppression of chaotic transport.   

Our approach is based on our recent work in 
Ref.~\cite{Fonseca14} where we proposed the gyro-averaged standard map (GSM) which generalizes the standard map by introducing the FLR dependence through the gyro-averaging of the drift-wave electrostatic potential. 
Here we focus on a statistical description of this problem. In particular, going beyond Ref.~\cite{Fonseca14} where all the plasma particles were assumed to have the same Larmor radius, $\hat{\rho}^0$, here we address the more realistic situation in which each of the $N$ particles of the plasma has a distinct Larmor radius,  $\hat{\rho}^i$, (with $i=1,\, \ldots N$) which is treated as a random variable sampled from a Maxwellian probability density function (pdf).  
As a result, the constant parameter $K=K_{0}J_{0}\left(\hat{\rho}^0\right)$  in the GSM, which corresponds to the effective drift wave amplitude with $J_0$ denoting the zeroth-order Bessel function, becomes the random variable $K=K_{0}J_{0}\left(\hat{\rho}^i\right)$.
That is, each particle has its ``own" GSM with effective drift-wave amplitude
$K(\hat{\rho}^i)$ and  the evolution of  the system is determined from the statistics of an ensemble of GSM maps.

The rest of the paper is organized as follows. Section \ref{sec:model} presents the transport model along with a brief review of the GSM. 
Starting from a Maxwellian pdf of Larmor radii, in Sec.~\ref{sec:Ypdf} we derive an analytical expression for the 
pdf of the effective drift wave amplitude $K=K_{0}J_{0}\left(\hat{\rho}\right)$, and compute the statistical moments and corresponding cumulative distribution function. The analytical results are compared with Monte-Carlo direct numerical simulations. 
Based on these results, Sec.~\ref{sec:SC} presents a statistical study of the confinement properties of the system. In particular, the probability of global chaos, $P_c$, (i.e., the probability  that a given plasma particle could in principle not be confined) is analytically computed and the results compared with  Monte-Carlo direct numerical simulations of the escape rate
for different values of $K_0$ and the thermal Larmor radius.
The results show that $P_c$ is an upper bound for the escape rate $\eta_e$. 
  Section~\ref{sec:Rt} studies the statistics of particle trapping in the plasma drift-wave main resonance. The
  probability of trapping, $P_t$, and the rate of trapping, $\eta_t$, are computed numerically  and compared
  for different values of $K_0$ and the thermal Larmor radius. 
The conclusions, including a summary of the results are presented in section \ref{sec:conclusion}.  

\section{Transport model}\label{sec:model}

In this section we present a brief summary of the Gyro-averaged Standard Map (GSM) model originally discussed in 
Ref.~\cite{Fonseca14}. 
The starting point is the $\vec{E}\times\vec{B}$ drift velocity of the guiding center \cite{white}
\begin{equation}
   \vec{V}_{E} = \frac{\vec{E}\times\vec{B}}{B^{2}} \, ,
   \label{eq:cg1}
\end{equation}
where $\vec{E}$ is the electric field, and $\vec{B}$ is the magnetic field with magnitude  $B=|\vec{B}|$.
Denoting with $x$  and $y$ the radial and poloidal coordinates  and  
writing $\vec{V}_{E} = (\dot{x}(t),\dot{y}(t))$ we get from Eq.~(\ref{eq:cg1}) the  $\vec{E}\times\vec{B}$ drift equations of motion   \begin{align}
     \frac{dx}{dt}=-\frac{1}{B_{0}}\frac{\partial \phi}{\partial y}, \qquad 
     \frac{dy}{dt}=\frac{1}{B_{0}}\frac{\partial \phi}{\partial x}
     \label{eq:system}
   \end{align}
where $\phi$ is the electrostatic potential and $B_{0}$ is the magnitude of the constant toroidal magnetic field.

Following Ref.~\cite{Lee}, we incorporate 
Finite Larmor radius (FLR) effects  by averaging the
electrostatic potential over a circle around the guiding center,
\begin{align}
   \langle\phi\rangle_{\varphi} &=\frac{1}{2\pi}\int_{0}^{2\pi}\phi(x+\rho\cos\varphi,\, y+\rho\sin\varphi, t)\, d\varphi \, ,\label{eq:FlrAverage}
 \end{align}
where $\rho$ is the Larmor radius.
Applying the gyro-averaging  $\langle ... \rangle_{\varphi}$  to Eq.~(\ref{eq:system}), we get the gyro-averaged  $\vec{E}\times\vec{B}$ drift equations of motion 
\begin{align}
   \frac{dx}{dt}=-\frac{1}{B_{0}}\frac{\partial\langle \phi \rangle_{\varphi}}{\partial y}, \qquad 
   \frac{dy}{dt}=\frac{1}{B_{0}}\frac{\partial\langle \phi \rangle_{\varphi}}{\partial x} \, .
   \label{eq:gyrosystem}
\end{align}

As a simplified model of weak drift-wave turbulence, following \cite{Horton98,Fonseca14}, we assume an electrostatic potential of the form
\begin{align}
   \phi\left(x,y,t\right)=\phi_{0}(x)+\sum_{m=-\infty}^{+\infty}A\cos(ky-m\omega_{0}t),
   \label{gyro_amplitude}
\end{align} 
where $\phi_{0}(x)$ is the equilibrium potential, $A$, is the amplitude of the drift waves, $k$ is the wave number, and 
$\omega_{0}$ is a fundamental frequency. 
The  corresponding gyro-averaged electrostatic potential is then given by, 
\begin{equation}
   \langle \phi(x,y,t)\rangle_{\varphi}=\langle \phi_{0}(x)\rangle_{\varphi}+AJ_{0}(\hat{\rho})\sum_{m=-\infty}^{+\infty}\cos\left(ky-m\omega_{0}t\right),
   \label{eq:gyroham}
\end{equation}
where $J_0$ is the zero-order Bessel function and $\hat{\rho}= k\rho$ is the normalized Larmor radius. For the sake of
brevity, from now on, we will refer to $\hat{\rho} $ as  the ``Larmor radius".
Using the Fourier series representation of the Dirac delta function, Eq.~(\ref{eq:gyroham}) can be written as
\begin{equation}
   \langle \phi(x,y,t)\rangle_{\varphi}=\langle \phi_{0}(x)\rangle_{\varphi} + 2\pi A J_{0}(\hat{\rho})\cos(ky)\sum_{m=-\infty}^{+\infty}\delta(\omega_{0}t-2\pi m).
   \label{eq:gyrohamdelta}
\end{equation}

Let  $x_n  = x(t^{-}_n)$ and $y_n  = y(t^{-}_n)$, with $t^{-}_n = \frac{2\pi n}{\omega_{0}}- \varepsilon$,  $n \in \mathbb{N}$, and $\varepsilon \rightarrow 0^{+}$.
Integrating equations (\ref{eq:gyrosystem}) over the time interval $(t^{-}_n,t^{-}_{n+1})$, with  $\langle \phi \rangle_{\varphi}$ in  Eq.~(\ref{eq:gyrohamdelta}), we obtain the  discrete model  
\begin{align} 
   x_{n+1} = x_n+\frac{2\pi kA}{\omega_0 B_{0}}J_{0}\left(\hat{\rho}\right)\sin(ky_n), \qquad  
   y_{n+1} = y_n+\frac{2\pi}{\omega_0 B_{0}}\frac{d \langle \phi_{0}\rangle_{\theta}}{dx}\bigg|_{x=x_{n+1}} 
   \label{eq:dwmap}
\end{align}
Note that in order to preserve the Hamiltonian structure, the equation for $y^{n+1}$ is implicit. This ensures that the transformation $(x_n,y_n)\rightarrow (x_{n+1},y_{n+1})$ is  symplectic, i.e. area preserving in the present $2$-dimensional case. 
In the case,
$\phi_{0} \sim x^2$,  Eqs.~(\ref{eq:dwmap}) reduce to the \emph{gyro-averaged standard map} (GSM):
\begin{align}
   I_{n+1} = I_{n}+K(\hat{\rho})\sin\theta_{n}, \qquad
   \theta_{n+1} = \theta_{n}+I_{n+1},\quad mod\quad2\pi\label{eq:mapTheta-norm-2}
\end{align}
where $I_{n}$ and $\theta_{n}$ are non-dimensional variables proportional to $x_{n}$ and $y_{n}$, respectively
and
\begin{equation}
   K=K_{0}J_{0}\left(\hat{\rho}\right), \label{eq:effective-perturb}
\end{equation}
where $K_0$ is a constant. In this paper we focused in this case. However, it should bear in mind that the model in Eq.~(\ref{eq:dwmap}) is quite general and admits other interesting possibilities. For example, as discussed in Ref.~\cite{Fonseca14}, in the case of non-monotonic ${\bf E} \times {\bf B}$ shear flows, $\phi_{0} \sim a x^2 -b x^3$ (with $a$ and $b$ constant) and the model reduces to the   \emph{gyro-averaged standard non-twist map} that exhibits different (compared to the GSM) and very interesting transport properties. 
  
Note that although the GSM has the same structure as the well-known Chirikov-Taylor standard map \cite{taylor, chirikov79}, there is a key difference due to dependence of the perturbation parameter $K$ on the Larmor radius.
To explain this subtle and crucial difference, consider a plasma consisting of an ensemble of $N$ particles
with Larmor radii $\hat{\rho}^i$ which at time $n$ are located at $(I_n^i, \theta_n^i)$ with $i=1,\ldots N$. Then, according to the GSM the time evolution of the system is  governed by \begin{align}
   I^i_{n+1} = I^i_{n}+K(\hat{\rho^i})\sin\theta^i_{n}, \qquad
   \theta^i_{n+1} = \theta^i_{n}+I^i_{n+1}  \, .\label{eq:mapTheta-norm-2_2} 
\end{align}
In the trivial and unrealistic case in which all the particles in the system have the same Larmor radius, say 
$\hat{\rho}^i=\hat{\rho}^0$ for $i=1,\,\ldots N$, Eq.~(\ref{eq:mapTheta-norm-2_2}) reduces to $N$-identical copies of a standard map with perturbation parameter $K=K_{0}J_{0}\left(\hat{\rho}^0\right)$, and the evolution of the system is well-understood. However, in a realistic plasma, collisional effects render the distribution of Larmor radii random. In this case, each particle has its ``own" standard-map like evolution with a perturbation parameter $K=K_{0}J_{0}\left(\hat{\rho}^i\right)$ (which is also a random variable) and  
the evolution of  the system is determined by the statistics of the ensemble of maps in Eq.~(\ref{eq:mapTheta-norm-2_2}).

\begin{figure}
   \begin{centering}
      \includegraphics[width=0.5\textwidth]{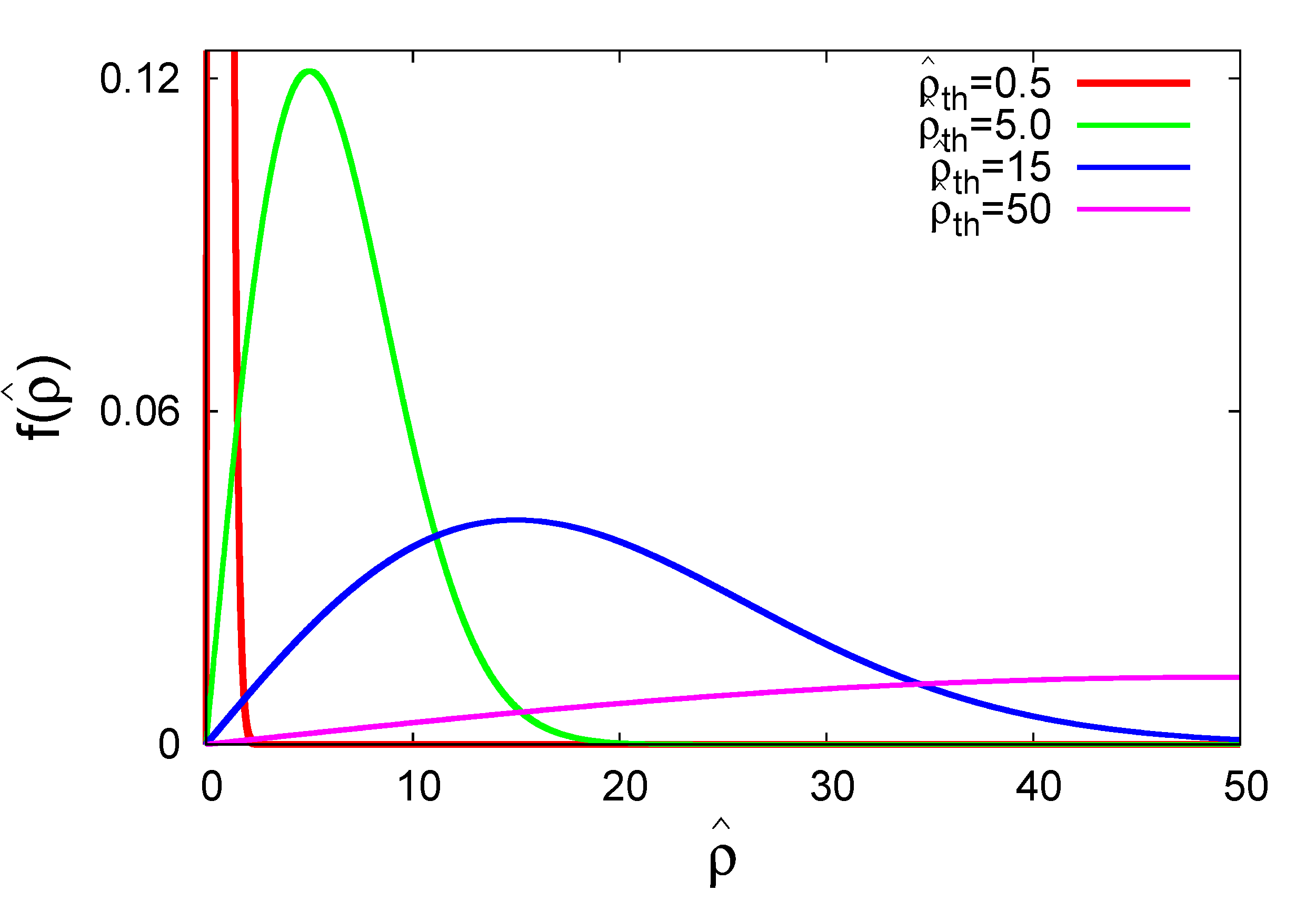}
   \par\end{centering}
   \caption{Larmor radius' probability density function of $\hat{\rho}$,  Eq. (\ref{eq:X-PDF}), for different values of the thermal Larmor radius, $\hat{\rho}_{th}$.}
   \label{fig:X-PDF}
\end{figure}

The previous discussion indicates that the GSM description of a plasma requires a model for the probability density function (pdf) of the Larmor radii of the particles, $f=f(\hat{\rho})$. 
As a simple realistic model, in this paper we assume that the plasma is in  thermal equilibrium which implies a
Maxwellian pdf of the form
\begin{equation}
    f\left(\hat{\rho}\right)=\frac{\hat{\rho}}{\hat{\rho}_{th}^{2}}\exp\left[-\frac{1}{2}\left(\frac{\hat{\rho}}{\hat{\rho}_{th}}\right)^{2}\right] 
    \label{eq:X-PDF}
\end{equation}
where 
\begin{equation}
   \hat{\rho}_{th}=\frac{k}{\left|q\right|B_{0}}\sqrt{mk_{B}T}
\end{equation}  
denotes the thermal Larmor radius, $q$ is the particle's charge, $m$  the particle's mass, $k_{B}$ is the Boltzmann constant, and $T$  is the plasma temperature.
Figure \ref{fig:X-PDF} shows plots of (\ref{eq:X-PDF}) for different values of the thermal Larmor radius.
The pdf has a maximum at  the most probable Larmor radius, $\hat{\rho}_{th}$, has 
mean value $ \left\langle \hat{\rho}\right\rangle = \sqrt{\frac{\pi}{2}}\hat{\rho}_{th}$, and 
variance $\sigma^{2} = \left\langle \hat{\rho}^{2}\right\rangle -\left\langle \hat{\rho}\right\rangle ^{2}=(4-\pi){\hat{\rho}_{th}}^2/2$.

\section{Statistics of gyro-averaged drift-wave amplitude}\label{sec:Ypdf} 

Given a model of the Larmor radius pdf, the statistical mechanics of the plasma in the GSM model is fundamentally controlled by the statistics of $\gamma=J_0(\hat{\rho})$ which according to Eq.~(\ref{eq:effective-perturb}) determines the gyro-averaged drift-wave perturbation effective amplitude. Therefore, as a first step in this section we derive the pdf of $\gamma=J_0(\hat{\rho})$ for the thermal equilibrium case in Eq.~(\ref{eq:X-PDF}).

\subsection{Probability density function}

According to  the random variable transformation theorem \cite{Gillepsie}, given  
the pdf of $\hat{\rho}$,  $f(\hat{\rho})$, the pdf of $\gamma$ is
\begin{equation}
   g(\gamma)=\int_{0}^{\infty}\delta\left[\gamma-J_{0}(\hat{\rho})\right]f(\hat{\rho})d\hat{\rho}\, .\label{eq:PDF_Y-RVT}
\end{equation}
Let $\Gamma_{\gamma}=\{\hat{\rho}_{0},\hat{\rho}_{1},\hat{\rho}_{2},...\}$
be the set of non-negative solutions of the equation
$\gamma=J_{0}(\hat{\rho}_{i})$ such that $J'_{0}(\hat{\rho}_{i})\neq0$, where the prime denotes the derivative.
If $\Gamma_{\gamma}$ is a non-empty set, the Dirac delta function
in (\ref{eq:PDF_Y-RVT}) can be rewritten as \cite{Kanwal}
\begin{equation}
  \delta\left[\gamma-J_{0}(\hat{\rho})\right]=\sum_{\hat{\rho}_{i}\in\Gamma_\gamma}\frac{\delta(\hat{\rho}-\hat{\rho}_{i})}{\left|J'_{0}(\hat{\rho}_{i})\right|} \, .
  \label{eq:delta-bessel}
\end{equation}
Note that $J'_{0}(\hat{\rho})=-J_{1}(\hat{\rho})$, where $J_{1}$ is the first-order Bessel function.
Substituting (\ref{eq:delta-bessel}) in (\ref{eq:PDF_Y-RVT}), we have
\begin{equation}
   g(\gamma)=\frac{1}{\hat{\rho}_{th}^{2}}\sum_{\hat{\rho}_{i}\in\Gamma_{\gamma}}\frac{\hat{\rho}_{i}}{\left|J'_{0}(\hat{\rho}_{i})\right|}\exp\left[-\frac{1}{2}\left(\frac{\hat{\rho}_{i}}{\hat{\rho}_{th}}\right)^{2}\right],\qquad\gamma_{min}<\gamma<1
  \label{eq:PDF_Y}
\end{equation}
where $\gamma_{min} \approx -0.4$ is
the smallest minimum of $J_{0}$, which corresponds to the first non-trivial zero of $J_1$. For $\gamma< \gamma_{min}$ 
and $\gamma>1$,  $\Gamma_{\gamma}$ is an empty set and $g(\gamma)=0$. 
It is interesting to observe that  (\ref{eq:PDF_Y}) is mathematically similar to the pdf obtained in Ref.~\cite{gustafson} in the context of a physically different problem.
 
\begin{figure}
\begin{centering}
\includegraphics[width=0.45\textwidth]{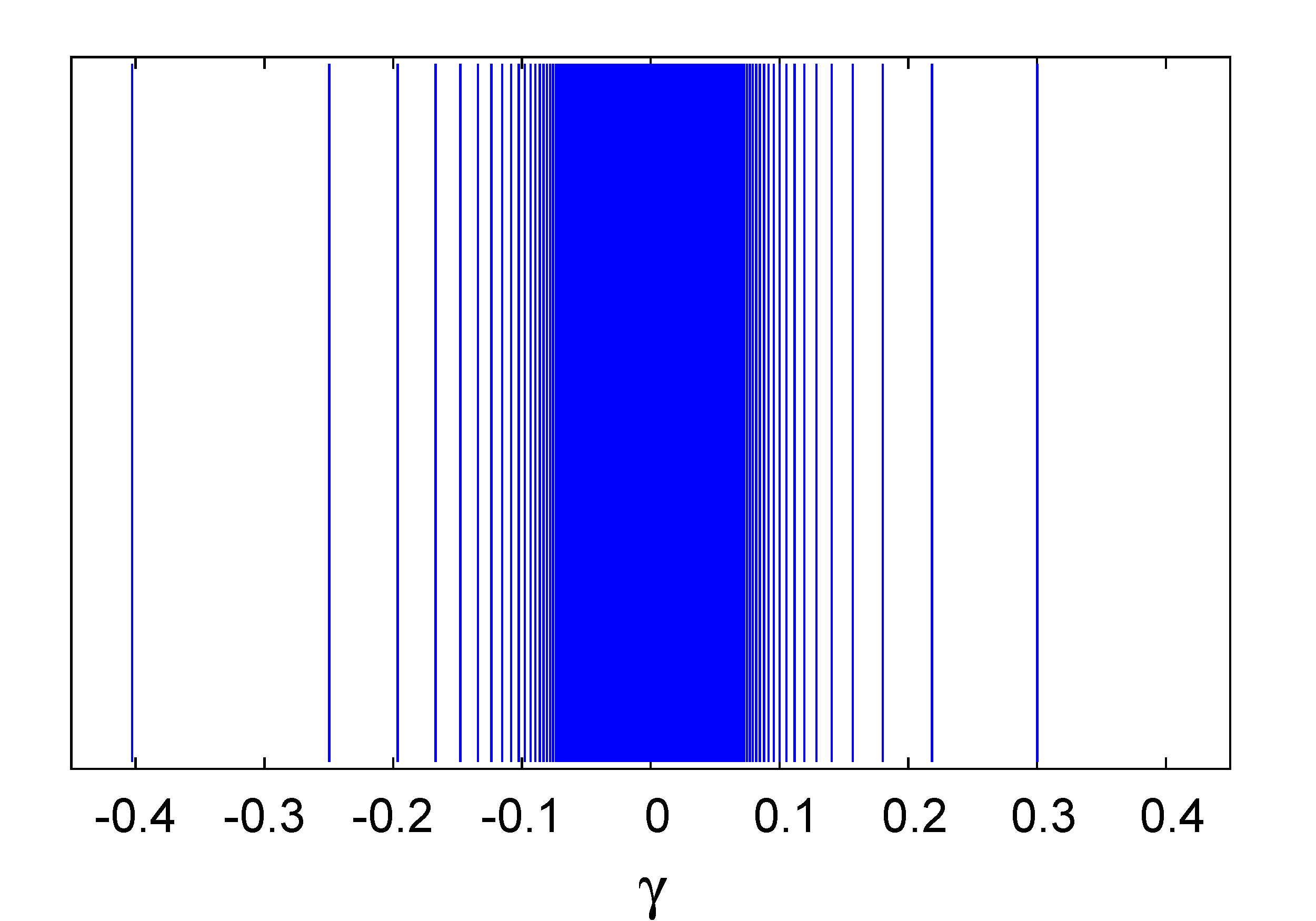}
\end{centering}
\caption{Vertical lines: location of singularities}
\label{fig:sing}
\end{figure}

The function $g$ consists of a sum of terms involving the product of 
$f(\hat{\rho})$, defined in (\ref{eq:X-PDF}), and the function $s(\hat{\rho})=1/\left|J'_{0}(\hat{\rho})\right|$,
 evaluated at values corresponding to the zeros of $\gamma=J_{0}(\hat{\rho})$. 
For values of $\gamma$ in the vicinity of a maximum or a minimum of $J_{0}$, $s$ diverges
That is, the minima and maxima of $J_{0}$ correspond to singularities of $g$.
As shown in Fig. \ref{fig:sing}, these singularities are not distributed homogeneously, and concentrate in the vicinity of $\gamma=0$. 
This property is a consequence of the asymptotic behavior of the zero-order Bessel function that, for large values of the argument, exhibits an oscillatory  decaying behavior of the form $J_{0}(\hat{\rho})\sim\sqrt{\frac{2}{\pi\hat{\rho}}}\cos(\hat{\rho}-\frac{\pi}{4})$. \cite{Abramowitz}. 

\begin{figure}
\begin{centering}
      \subfloat[$\hat{\rho}_{th}=0.5$]{\includegraphics[width=0.40\textwidth]{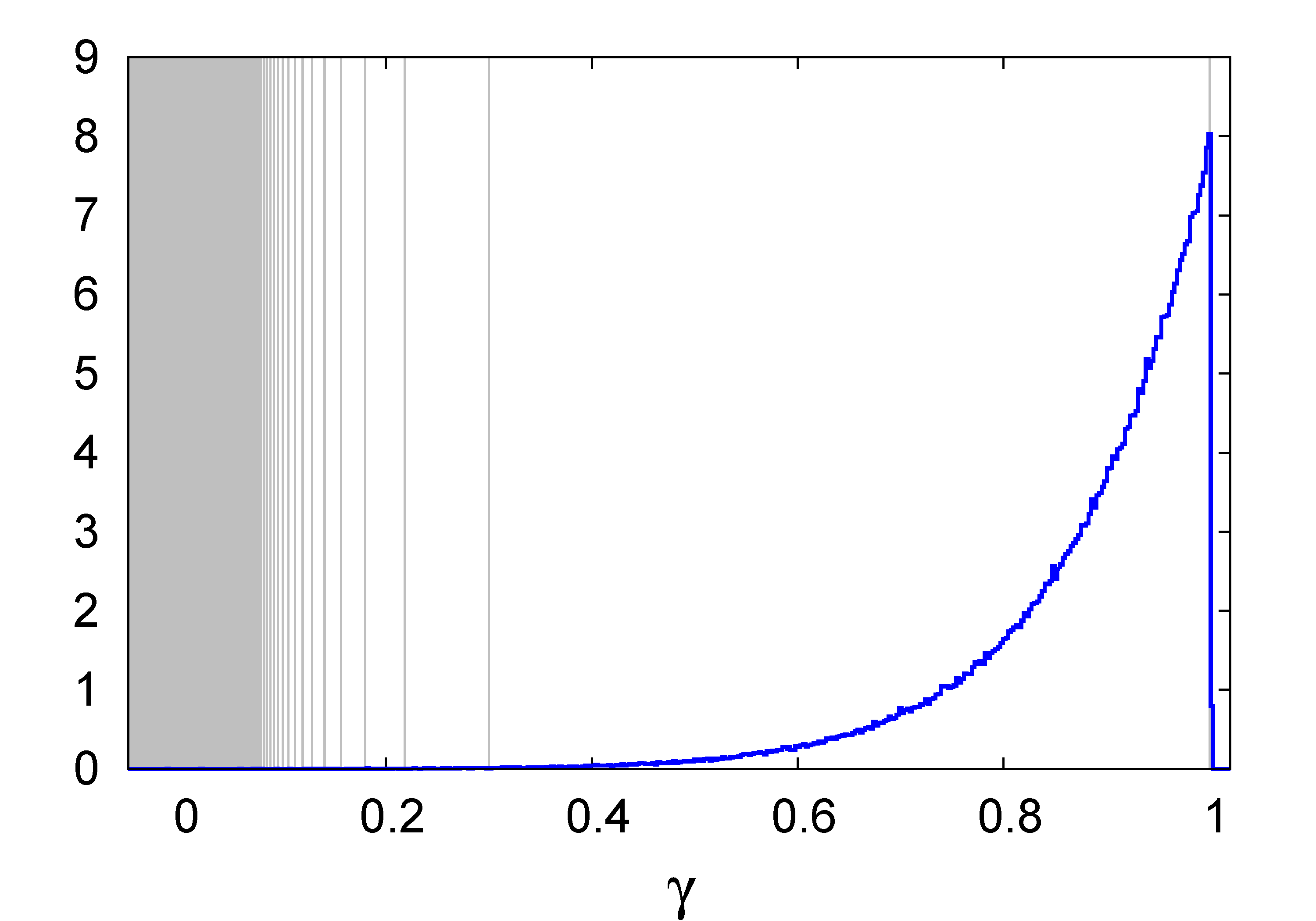}} ~ \subfloat[$\hat{\rho}_{th}=5$]{\includegraphics[width=0.40\textwidth]{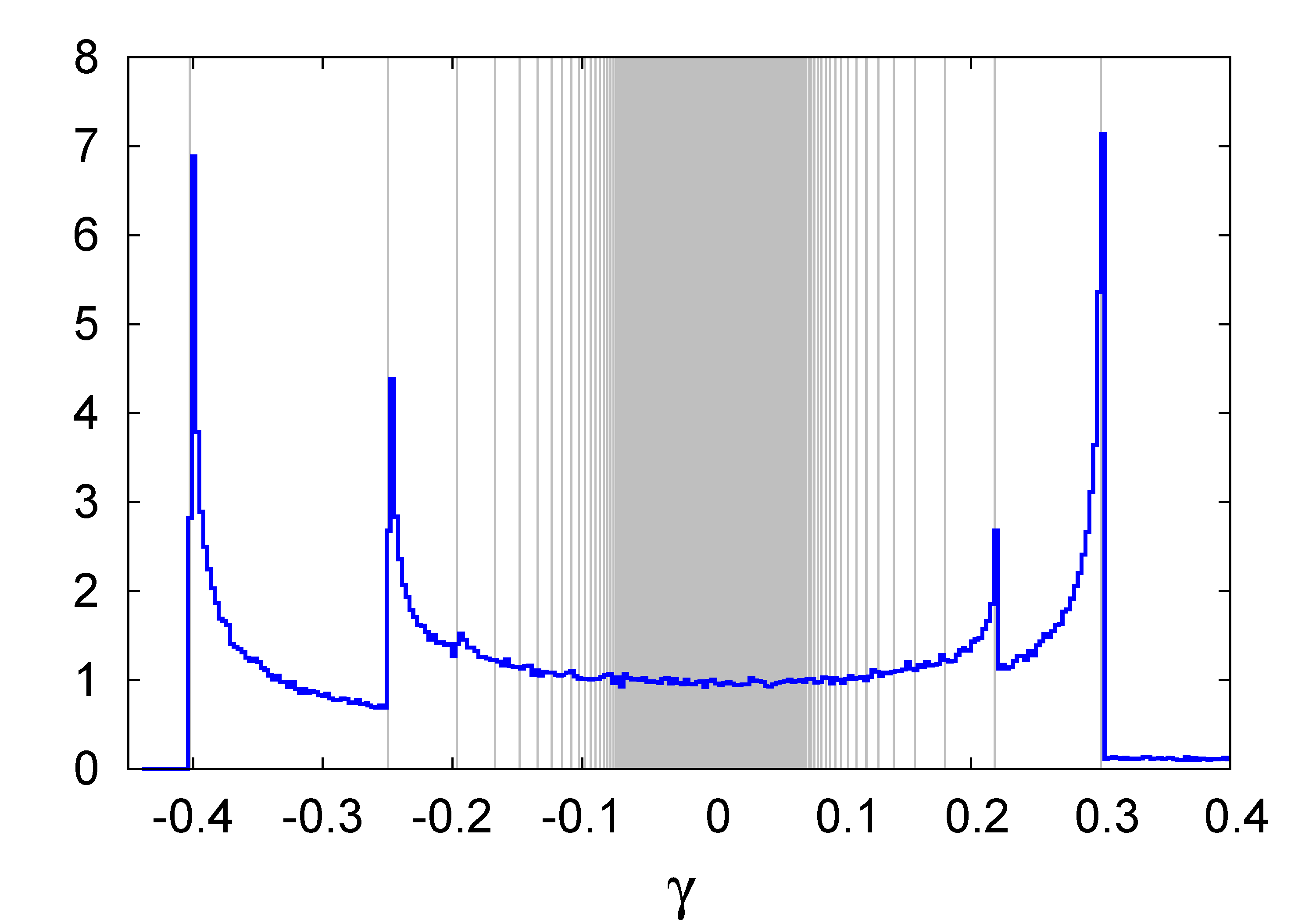}}
      \par\end{centering}
\begin{centering}      
      \subfloat[$\hat{\rho}_{th}=15$]{\includegraphics[width=0.40\textwidth]{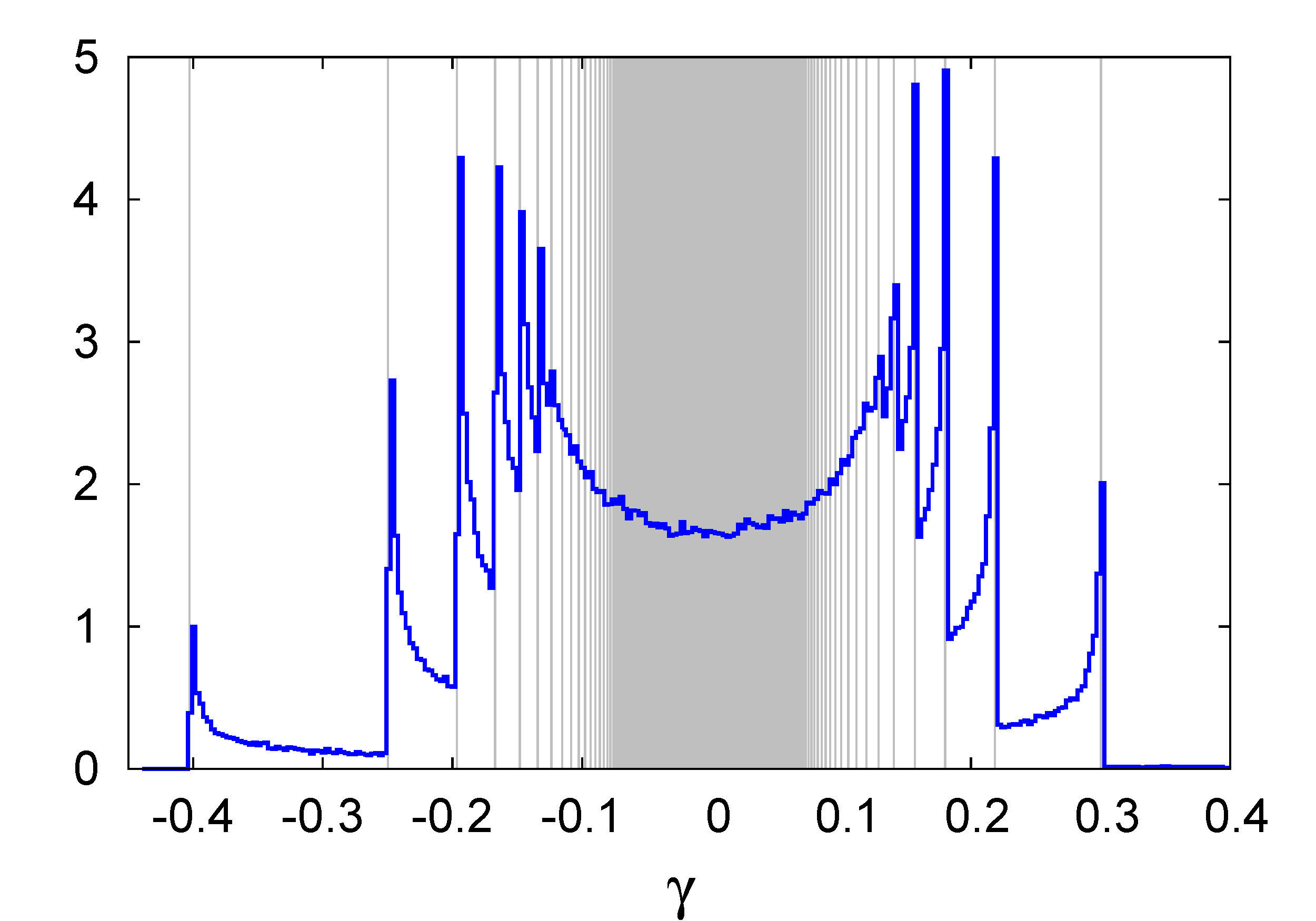}} ~ \subfloat[$\hat{\rho}_{th}=50$]{\includegraphics[width=0.40\textwidth]{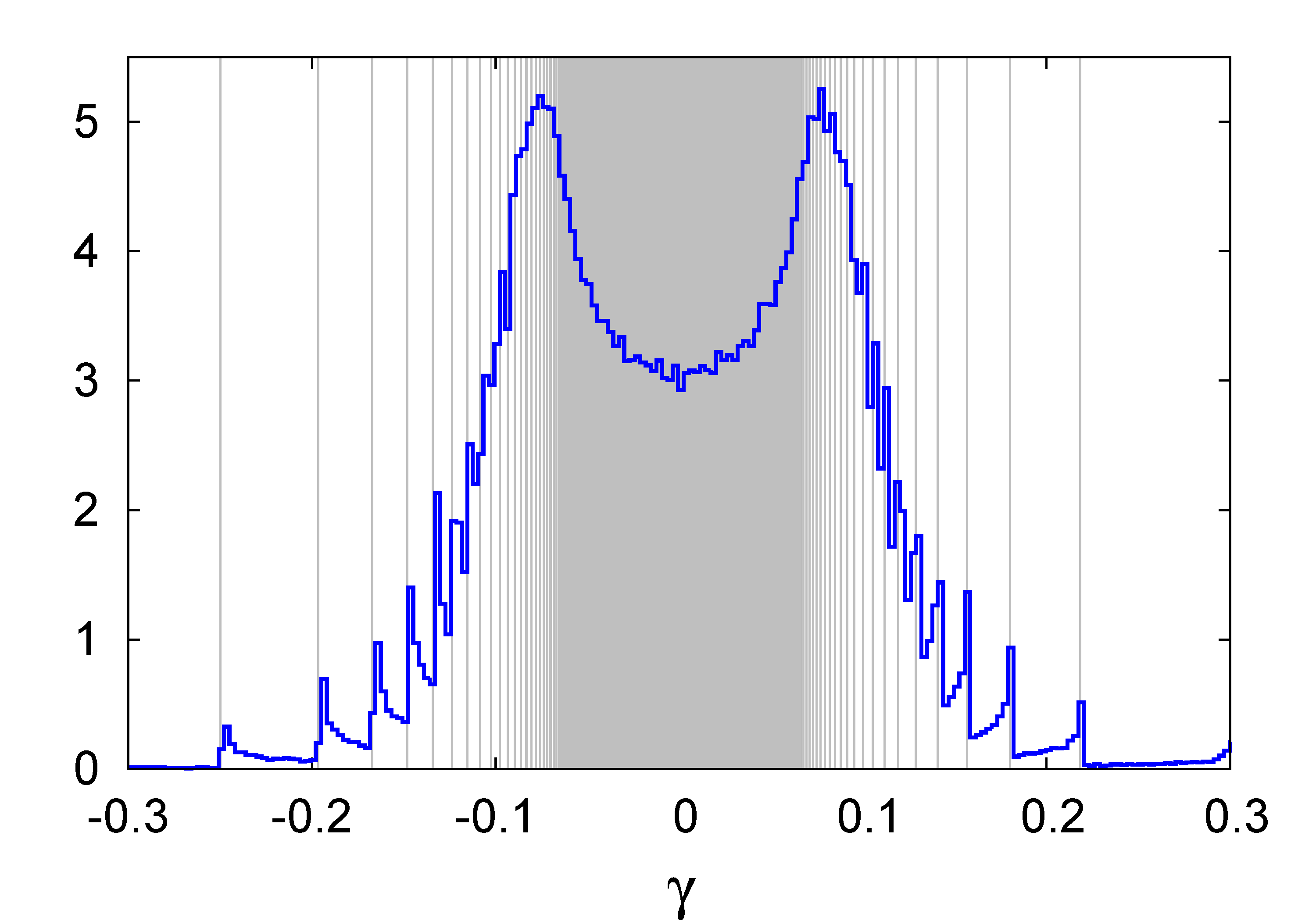}}
\par\end{centering}
\caption{Probability density function of the gyro-averaged drift-wave amplitude,$\gamma$, superimposed to the vertical lines indicating the singularities.}
\label{fig:hists}
\end{figure}

\begin{figure}
   \begin{centering}
      \subfloat[Function $f$ at the singularities and the pdf in Fig. \ref{fig:hists}(c).]{\includegraphics[width=0.45\textwidth]{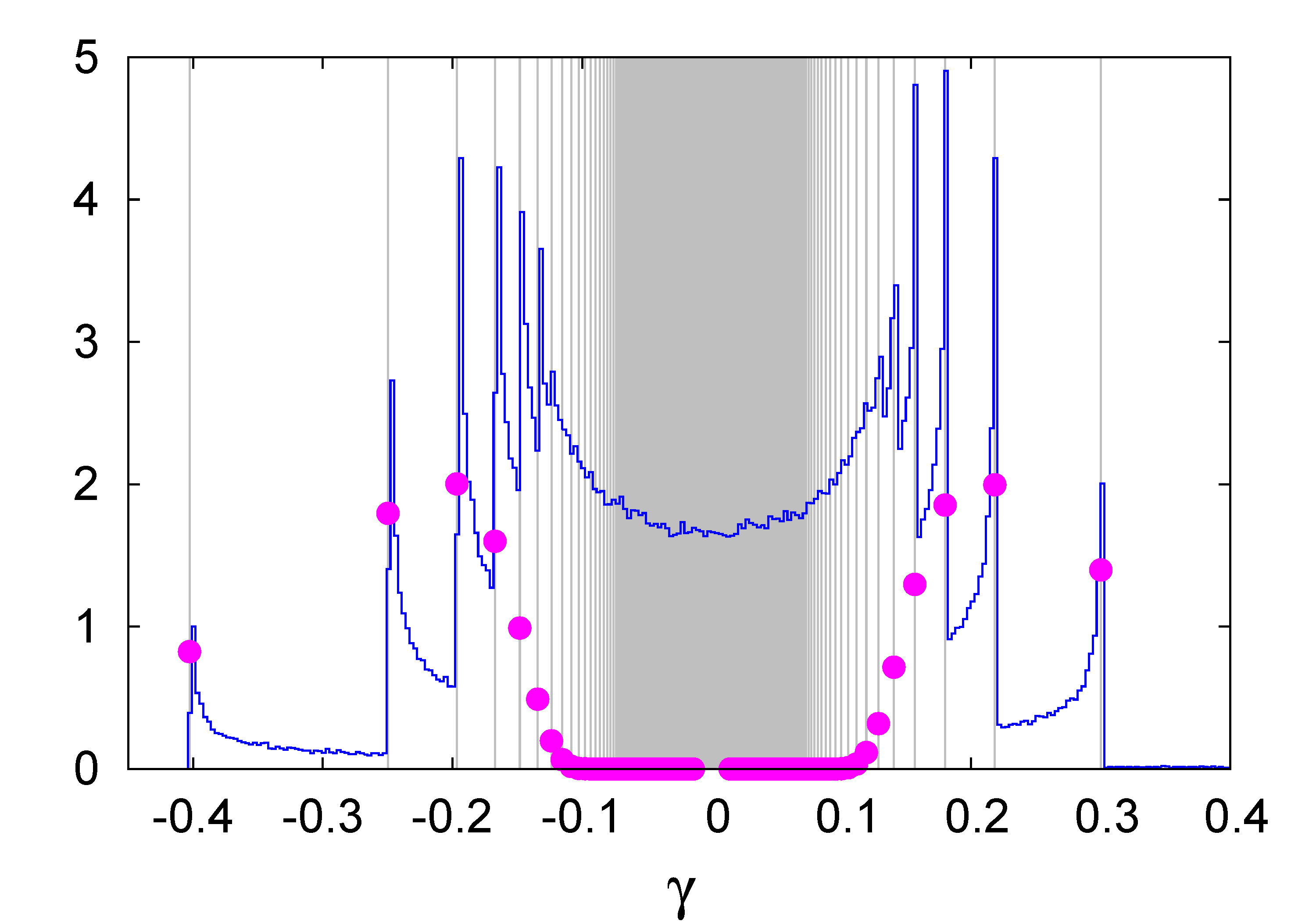}}
      ~\subfloat[Function $f$ at the singularities for different values of $\hat{\rho}_{th}$. ]{\includegraphics[width=0.45\textwidth]{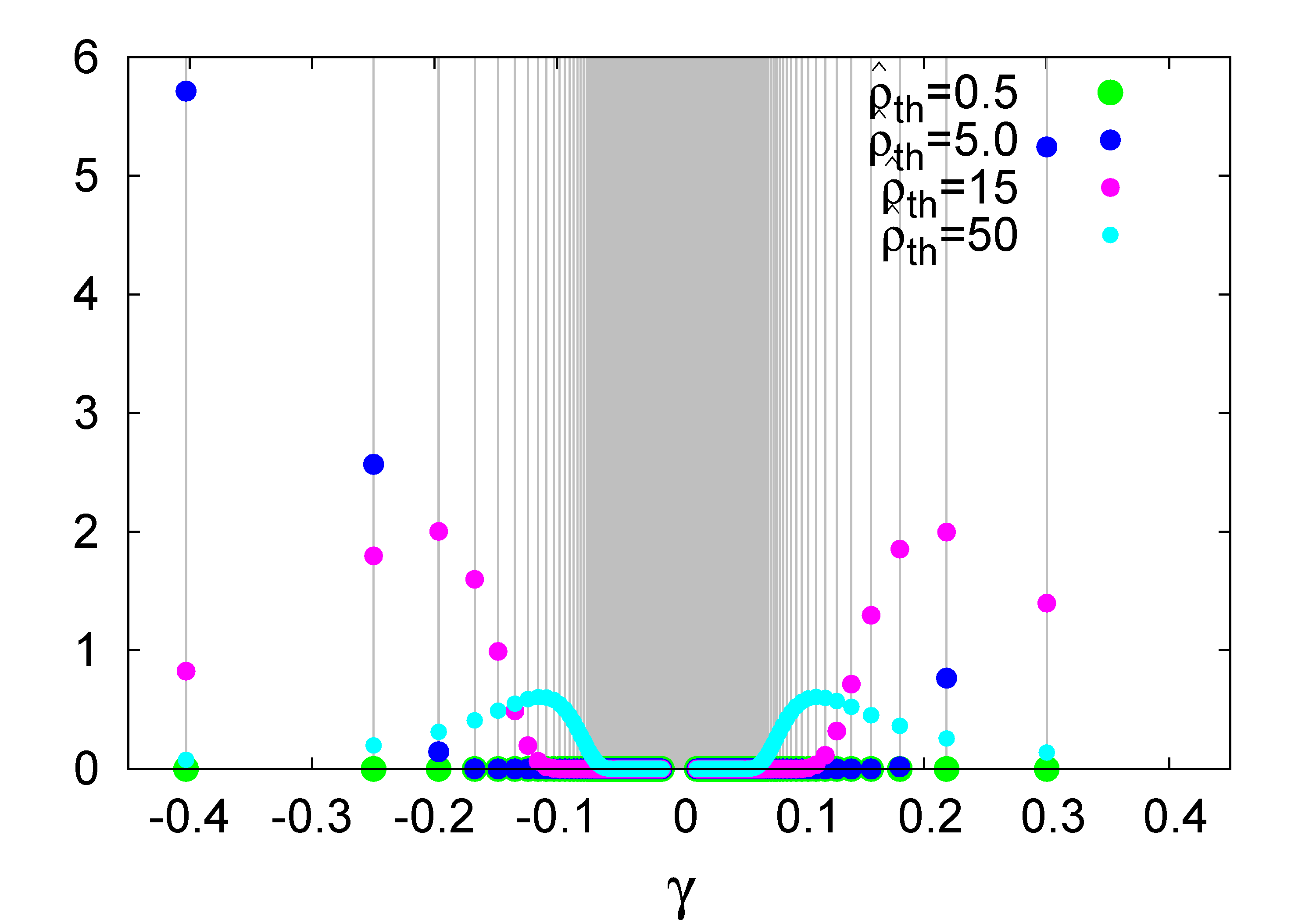}}
\end{centering}
\caption{Function $f$, defined by Eq.(\ref{eq:X-PDF}), evaluated at the points of singularity. The fast exponential decaying
character of  $f$ controls the effects of singularities,
acting as an ``height'' function and limiting the growth of $g(\gamma)$. 
For visualization purposes, the function $f$ (shown with dots) has been rescaled using a factor of $50$.
}
\label{fig:sing_height}
\end{figure}

To compute the $g$ in Eq.~(\ref{eq:PDF_Y}) we use a Monte Carlo method.  
First, an ensemble of random  Larmor radii is numerically generated from the pdf in  Eq. (\ref{eq:X-PDF}).
Then, for each random Larmor radius generated, the corresponding value of $\gamma$ is computed using the relation $\gamma=J_{0}(\hat{\rho})$, and a histogram of $\gamma$ values is constructed.  
Figure~\ref{fig:hists} shows the resulting pdf (obtained from the normalized histogram) for different values of the thermal Larmor radius $\hat{\rho}_{th}$. For reference, the figure also shows the location of the singularities of $g$ which correlate with the ``peaks'' of $g$. However, not all singularities are accompanied  by peaks in the pdf. We  will refer to a singularity as  
``strong'' if a  ``peak'' is observed in the pdf, and as ``weak'' if this is not the case. 
What controls the strength of a singularity is the exponential decay
of $f$ that limits the magnitud of $g(\gamma)$.
To clarify this,  Fig. \ref{fig:sing_height}(a) shows the pdf of $\gamma$ along with the 
function $f$ evaluated at the location of the singularities. As expected, the pdf of $\gamma$ exhibits peaks where $f$ has higher values and the singularities are more distant to each other.
The peaks are not observed near $\gamma=0$ where $f$ goes to zero
and the singularities are more concentrated. The dependence of these observations on the value of $\hat{\rho}_{th}$ is explored in Fig. \ref{fig:sing_height}(b) that shows  plots of $f$ evaluated at the singularities
for the different values of $\hat{\rho}_{th}$ used in the pdfs in Figs. \ref{fig:hists}(a)-(c). 
Once again,  it is observed that  peaks predominate at the singularities with the highest values of $f$. 

\subsection{Statistical moments}

The $n$-th moment of $\gamma$ is given by
\begin{equation}
   \left\langle \gamma^{n}\right\rangle =\intop_{\gamma_{min}}^{1}\gamma^{n}g(\gamma)d\gamma \, ,
   \label{eq:moment}
\end{equation}
where we have used the fact that, as explained in the previous section, $g(\gamma)=0$ for $\gamma<\gamma_{min}$ and 
$\gamma>1$. 
Substituting (\ref{eq:PDF_Y-RVT}) in (\ref{eq:moment}), we have:
\begin{equation}
   \left\langle \gamma^{n}\right\rangle =\frac{1}{\hat{\rho}_{th}^{2}}\intop_{0}^{\infty}\left[J_{0}(\hat{\rho})\right]^{n}\exp\left[-\frac{1}{2}\left(\frac{\hat{\rho}}{\hat{\rho}_{th}}\right)^{2}\right]\hat{\rho}d\hat{\rho}
   \label{eq:nth-moment}
\end{equation}
Using  the identity in Eq. (6.631-4) of Ref. \cite{Gradshteyn},
\begin{equation}
   \intop_{0}^{+\infty}\hat{\rho}^{\nu+1}e^{-\alpha \hat{\rho}^{2}}J_{\nu}(\beta \hat{\rho})d\hat{\rho}=\frac{\beta^{\nu}}{(2\alpha)^{\nu+1}}\exp\left(-\frac{\beta^{2}}{4\alpha}\right),
   \label{eq:rel}
\end{equation}
with $\alpha=\frac{1}{2\hat{\rho}_{th}^{2}}$,
$\beta=1$, and $\nu=0$, it follows that  
\begin{equation}
   \left\langle K \right\rangle = K_{0}  \left\langle \gamma\right\rangle = K_{0} \exp\left(-\frac{\hat{\rho}_{th}^{2}}{2}\right) \, .
   \label{eq:Kaverage}
\end{equation}
That is, the mean of the effective drift wave perturbation amplitude 
decreases exponentially with increasing values of the thermal Larmor radius. 

\begin{figure}
\begin{centering}
\includegraphics[width=0.45\textwidth]{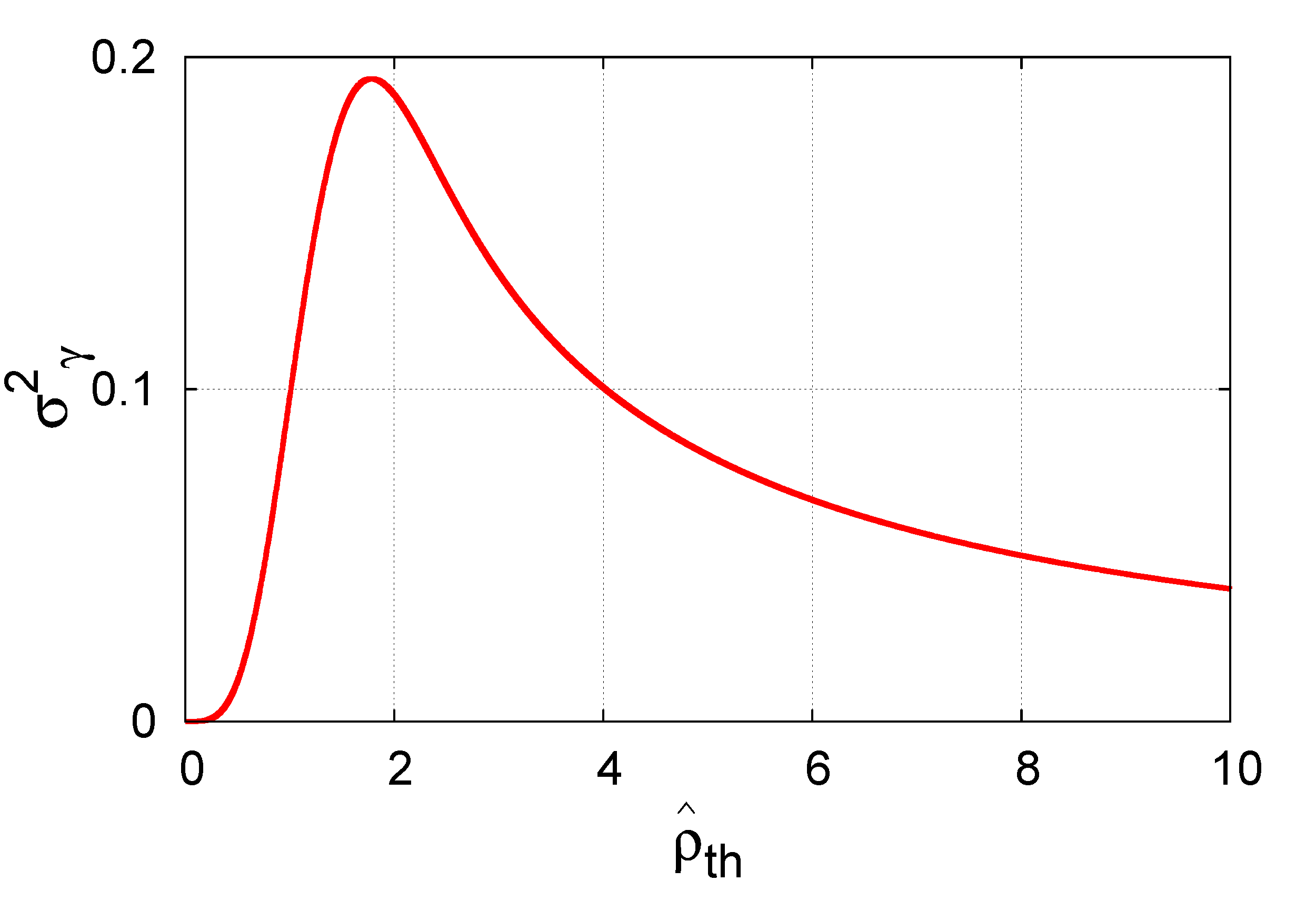}
\par\end{centering}
\caption{Dispersion of $\gamma$, $\sigma_{\gamma}^{2}$. For increasing $\hat{\rho}_{th}$,  $\sigma_{\gamma}^{2}$ increases
from zero to maximum and then goes to zero again at high values of $\hat{\rho}_{th}$.}
\label{fig:variance}
\end{figure}

For the second moment, $\left\langle \gamma^{2}\right\rangle$, we use Eq. (6.633-2) of Ref. \cite{Gradshteyn},  
\begin{equation}
   \intop_{0}^{+\infty}\hat{\rho}e^{-\varrho^{2}\hat{\rho}^{2}}J_{p}(\alpha \hat{\rho})J_{p}(\beta \hat{\rho})d\hat{\rho}=\frac{1}{2\varrho^{2}}\exp\left[\frac{\alpha^{2}+\beta^{2}}{4\varrho^{2}}\right]I_{p}\left(\frac{\alpha\beta}{2\varrho^{2}}\right)\, ,
   \label{eq:rel2}
\end{equation}
with   $\varrho^{2}=\frac{1}{2\hat{\rho}_{th}^{2}}$, $\alpha=\beta=1$, and $p=0$, and conclude
\begin{equation}
  \left\langle K^2 \right\rangle = K^2_{0}  \left\langle \gamma^2 \right\rangle = e^{-\hat{\rho}_{th}^{2}}I_{0}\left(\hat{\rho}_{th}^{2}\right) \, .
   \label{eq:second-moment}
\end{equation}
Finally, using (\ref{eq:Kaverage}) and (\ref{eq:second-moment}), 
the dispersion of the effective perturbation, defined as $\sigma_{K}^{2} = K_0^2\left(\left\langle \gamma^{2}\right\rangle - \left\langle \gamma\right\rangle^{2} \right)$ is  given by
\begin{align}
  \sigma_{K}^{2}=K_{0}^2e^{-\hat{\rho}_{th}^{2}}\left[I_{0}\left(\hat{\rho}_{th}^{2}\right)-1\right]
  \label{eq:Kdispersion}
\end{align}
As observed in Fig. \ref{fig:variance}, the dispersion increases
for small values of $\hat{\rho}_{th}$ from zero to a maximum
and then decays. For large $\hat{\rho}_{th}$, since $I_{0}\left(\hat{\rho}_{th}^{2}\right)\sim\frac{e^{+\hat{\rho}_{th}^{2}}}{\sqrt{2\pi\hat{\rho}_{th}^{2}}}$ \cite{Abramowitz},
$\sigma_{K}^{2}$ decays as $\sigma_{K}^{2}\sim\frac{K_0^2}{\hat{\rho}_{th}}$.
Thus, the dispersion ``expands'' from zero to a maximum and then ``compress''
to zero again for increasing $\hat{\rho}_{th}$.
The pdf in Figs. \ref{fig:hists}(a)-(d) also indicate that, for increasing $\hat{\rho}_{th}$, both the average and the dispersion go to zero
because the peaks become more symmetrically concentrated near $\gamma=0$. 

\subsection{Cumulative Distribution Function}\label{sec:Fy} 

To compute the cumulative distribution function of $\gamma$, 
\begin{equation}
   G\left(\gamma\right)=\int_{-\infty}^{\gamma}g(\gamma')d\gamma' \, ,
   \label{eq:Fy_Dist_Def}
\end{equation}
we substitute (\ref{eq:PDF_Y-RVT}) in (\ref{eq:Fy_Dist_Def}),
\begin{equation}
   G\left(\gamma\right)=\int_{0}^{\infty}\Theta[\gamma-J_{0}(\hat{\rho})]f(\hat{\rho})d\hat{\rho}\label{eq:Fy_Theta}
\end{equation}
where $\Theta$ is the Heaviside step function. Integrating by parts Eq.~(\ref{eq:Fy_Theta}) yields
\begin{equation}
   G\left(\gamma\right)=\Theta(\gamma-1)-\int_{0}^{\infty}\delta\left[\gamma-J_{0}(\hat{\rho})\right]J_{0}'(\hat{\rho})\exp\left[-\frac{1}{2}\left(\frac{\hat{\rho}}{\hat{\rho}_{th}}\right)^{2}\right]d\hat{\rho}
   \label{eq:FyPartialIntegration}
\end{equation}
If $\gamma$ is outside the interval $\gamma_{min}<\gamma<1$, the second term in (\ref{eq:FyPartialIntegration}) vanishes and
 $G\left(\gamma\right)=\Theta(\gamma-1)$.
That is, $G(\gamma)=0$ for $\gamma \leq \gamma_{min}$ and $G(\gamma)=1$ for $\gamma \geq 1$.
On the other hand, for $\gamma_{min}<\gamma<1$,  $\Theta(\gamma-1)=0$ and, using again formula (\ref{eq:delta-bessel}),
Eq.  (\ref{eq:FyPartialIntegration}) yields
\begin{equation}
   G\left(\gamma\right)=-\sum_{\hat{\rho}_{i}\in\Gamma_{\gamma}}\frac{J'_{0}(\hat{\rho}_{i})}{\left|J'_{0}(\hat{\rho}_{i})\right|} \exp\left[-\frac{1}{2}\left(\frac{\hat{\rho}_{i}}{\hat{\rho}_{th}}\right)^{2}\right] \, .
   \label{eq:Fy-withroots}
\end{equation}

A simpler form for (\ref{eq:Fy-withroots}) can be obtained if we define an additional property for $\Gamma_{\gamma}$. 
Let $\Gamma_{\gamma}$ be an order set such that $\hat{\rho}_{0} < \hat{\rho}_{1} < \hat{\rho}_{2} <...$. 
Each solution $\hat{\rho}_{i}$ belongs to an interval where $J_{0}(\hat{\rho})$ is increasing or decreasing. 
$J_{0}(\hat{\rho})$ oscillates such that $J'_{0}(\hat{\rho}_{i})/\left|J'_{0}(\hat{\rho}_{i})\right|=-1$ for $i=0,2,4,..$ and 
$J'_{0}(\hat{\rho}_{i})/\left|J'_{0}(\hat{\rho}_{i})\right|=+1$ for $i=1,3,5,..$, or:
\begin{equation}
   J'_{0}(\hat{\rho}_{i})/\left|J'_{0}(\hat{\rho}_{i})\right|=(-1)^{i-1}, \quad i=0,1,2,...
   \label{eq:Fy-termssignal}
\end{equation}
Substituting (\ref{eq:Fy-termssignal}) in (\ref{eq:Fy-withroots}), we have the following expression 
for the cumulative distribution function:
\begin{align}
   G\left(\gamma\right) & =\left\{ \begin{array}{cc}
                                          0, & \gamma\leq\gamma_{min}\\
                                          \sum_{\hat{\rho}_{i}\in\Gamma_{\gamma}}(-1)^{i}\exp\left[-\frac{1}{2}\left(\frac{\hat{\rho}_{i}}{\hat{\rho}_{th}}\right)^{2}\right], & \gamma_{min}<\gamma<1\\
                                          1 & \gamma\geq 1
                                       \end{array}\right.
   \label{eq:FyFinalForm-1}
\end{align}

\begin{figure}
\begin{centering}
   \subfloat[$\hat{\rho}_{th}=0.5$]{\includegraphics[width=0.45\textwidth]{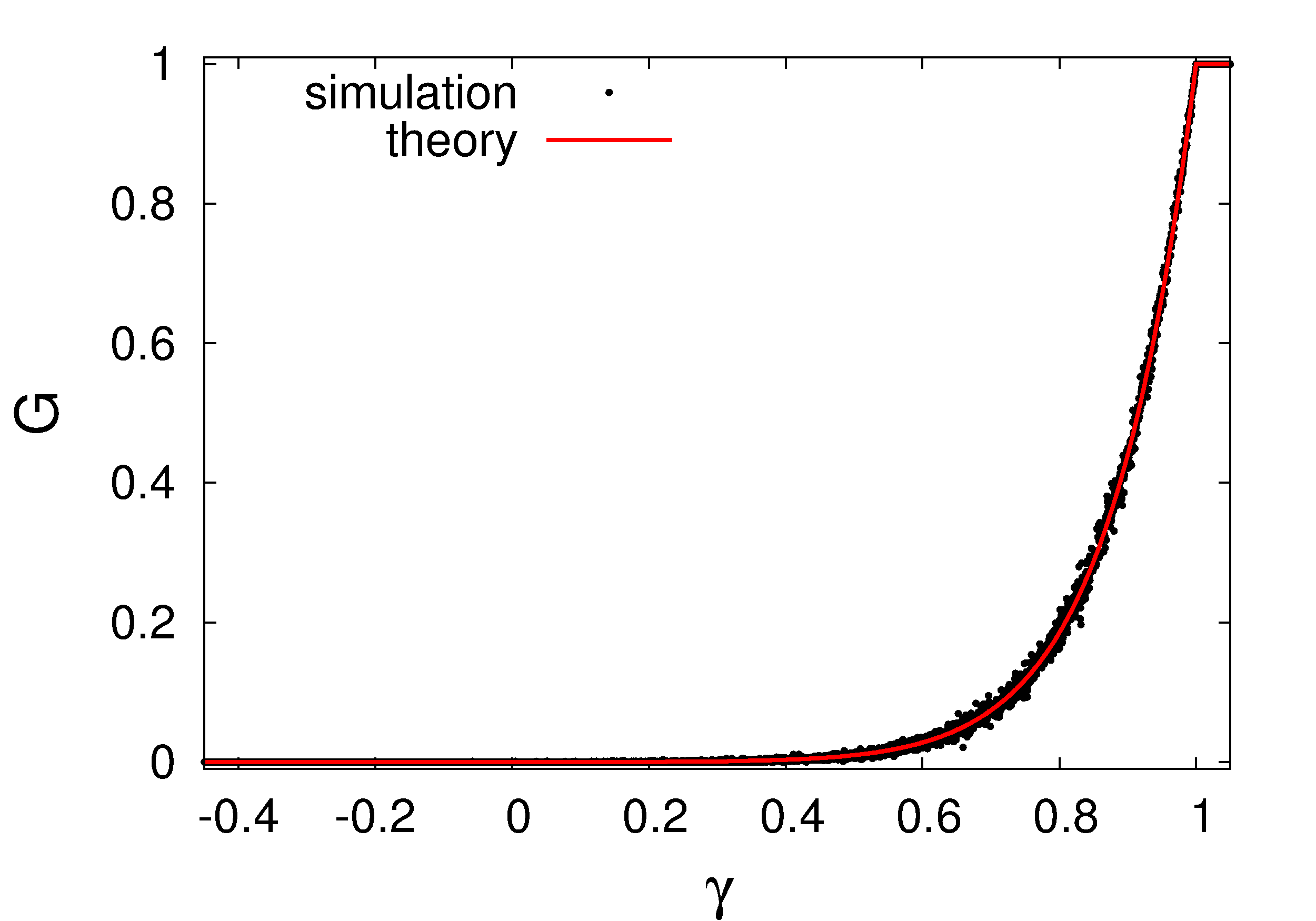}}
   ~\subfloat[$\hat{\rho}_{th}=5.0$]{\includegraphics[width=0.45\textwidth]{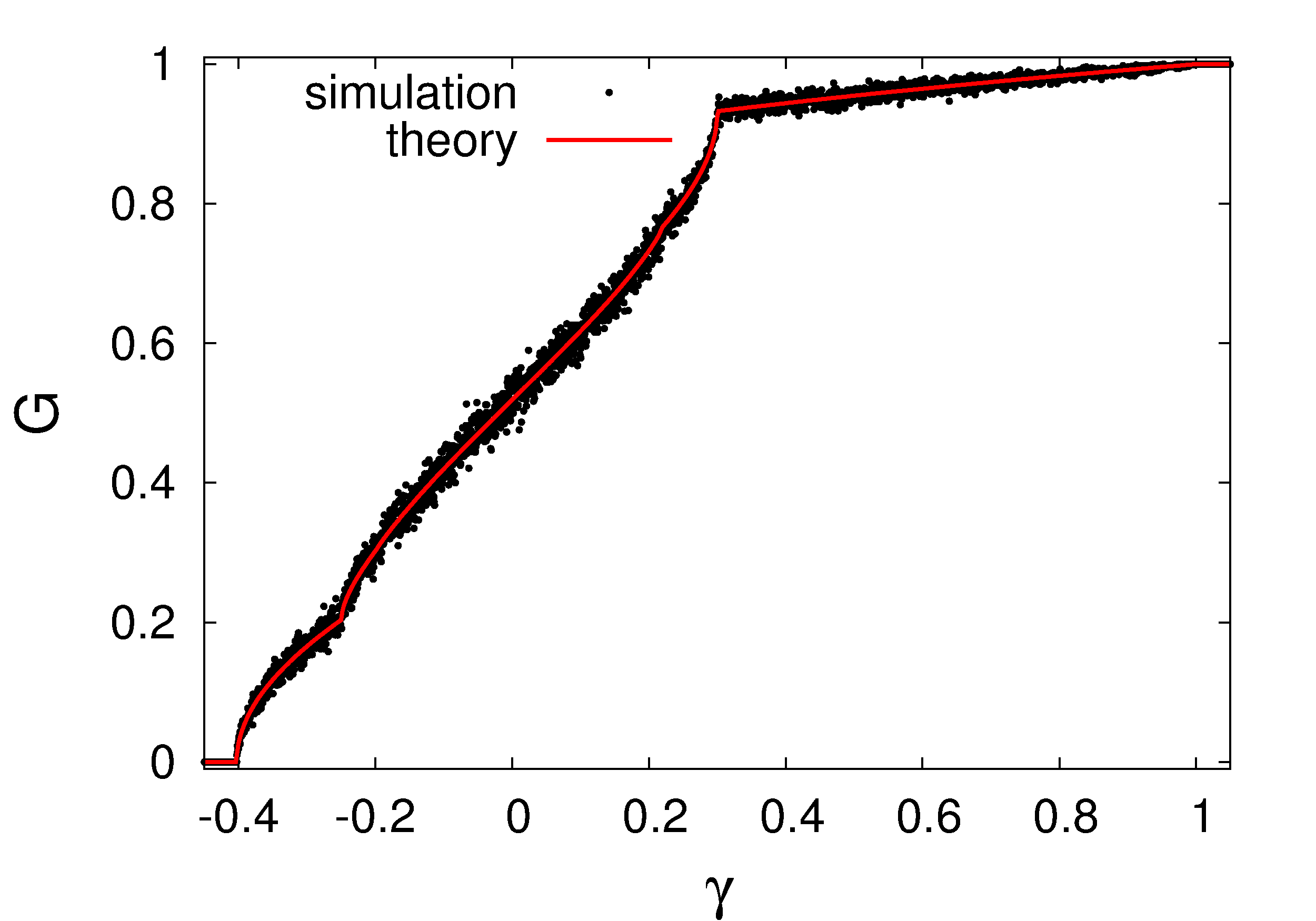}
}
\par\end{centering}
\begin{centering}
   \subfloat[$\hat{\rho}_{th}=15$]{\includegraphics[width=0.45\textwidth]{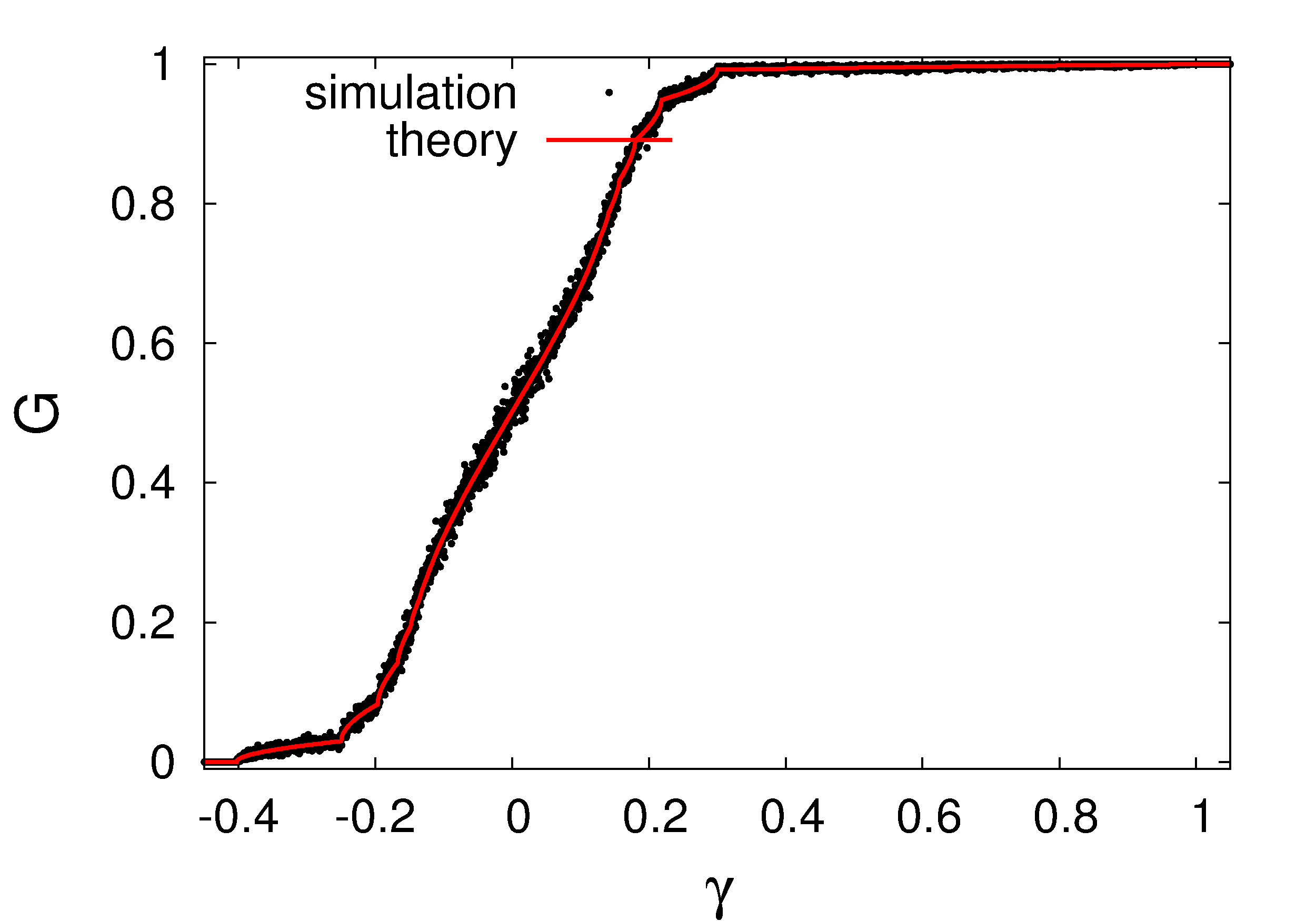}}
   ~\subfloat[$\hat{\rho}_{th}=50$]{\includegraphics[width=0.45\textwidth]{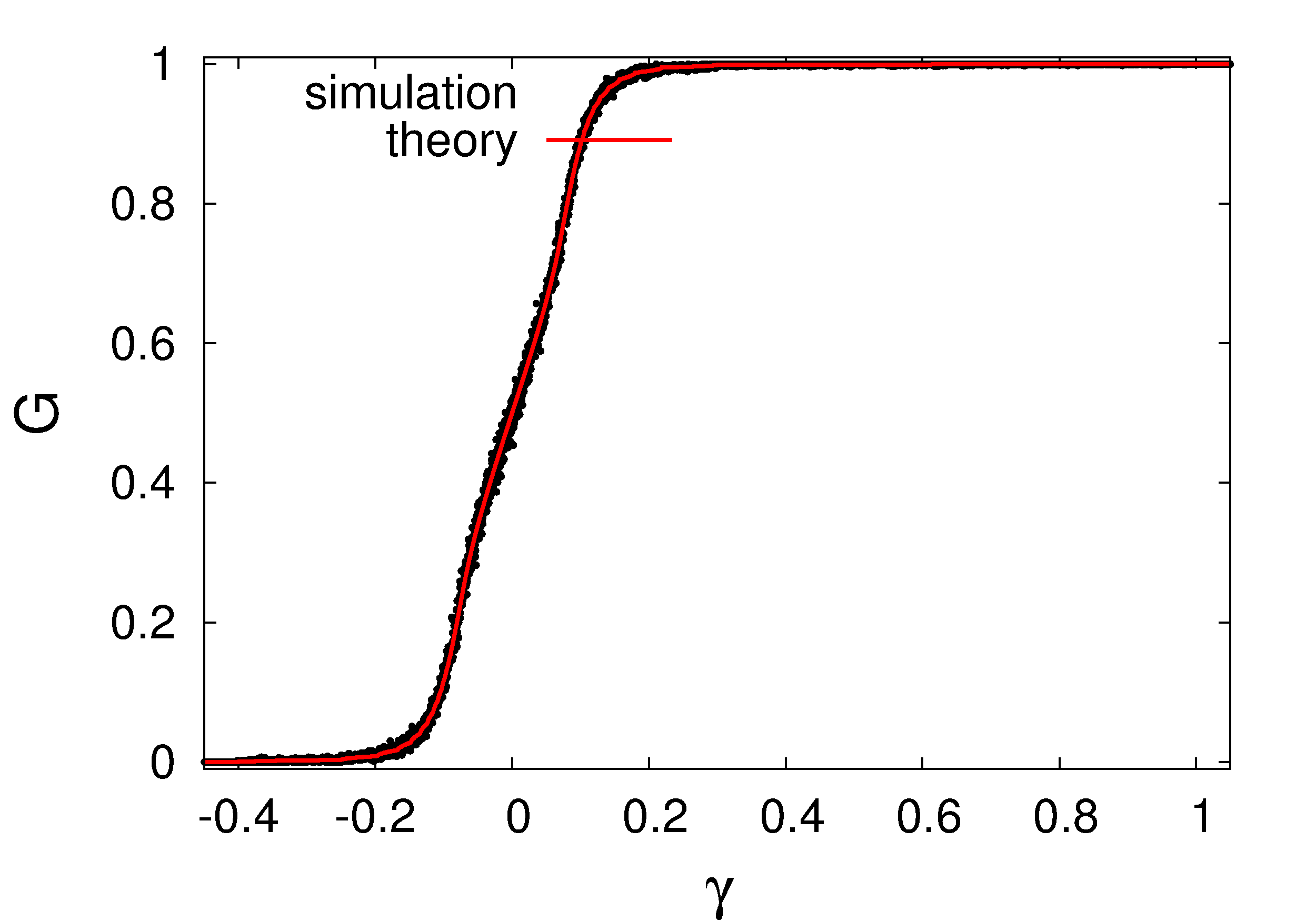}}
\par\end{centering}
\caption{Cumulative distribution function  of $\gamma$, denoted by $G(\gamma)$
The red curves are plots of the analytical result described by Eq.
(\ref{eq:FyFinalForm-1}). The black points correspond to numerical simulations.}
\label{fig:Fy}
\end{figure}

Figures \ref{fig:Fy}(a)-(b)
show very good agreement between the analytical result in Eq.~(\ref{eq:FyFinalForm-1}) (red curves) and 
Monte-Carlo  simulations of $G$ (black points)  for different values of $\hat{\rho}_{th}$. 
The Monte-Carlo simulations were performed as follows: first,  an ensemble of $N=1000$
random Larmor radii distributed according to $f$ was generated; second, for each Larmor radius value $\hat{\rho}$ generated,
we compute $J_{0}(\hat{\rho})$; finally, for a given $\gamma$, we determine
the rate or frequency of values $J_{0}(\hat{\rho})$ such that $J_{0}(\hat{\rho})\leq\gamma$.
As expected, since no value of $J_{0}$ can be below
$\gamma_{min}$, it is observed that  $G\left(\gamma\right) \rightarrow 0$  as $\gamma \rightarrow \gamma_{min}$. Also, 
since no values of $J_{0}$ can be above $\gamma=1$, $G\left(\gamma\right) \rightarrow 1$ as $\gamma \rightarrow 1$.

\begin{figure}
\begin{centering}
\subfloat[]{\includegraphics[width=0.45\textwidth]{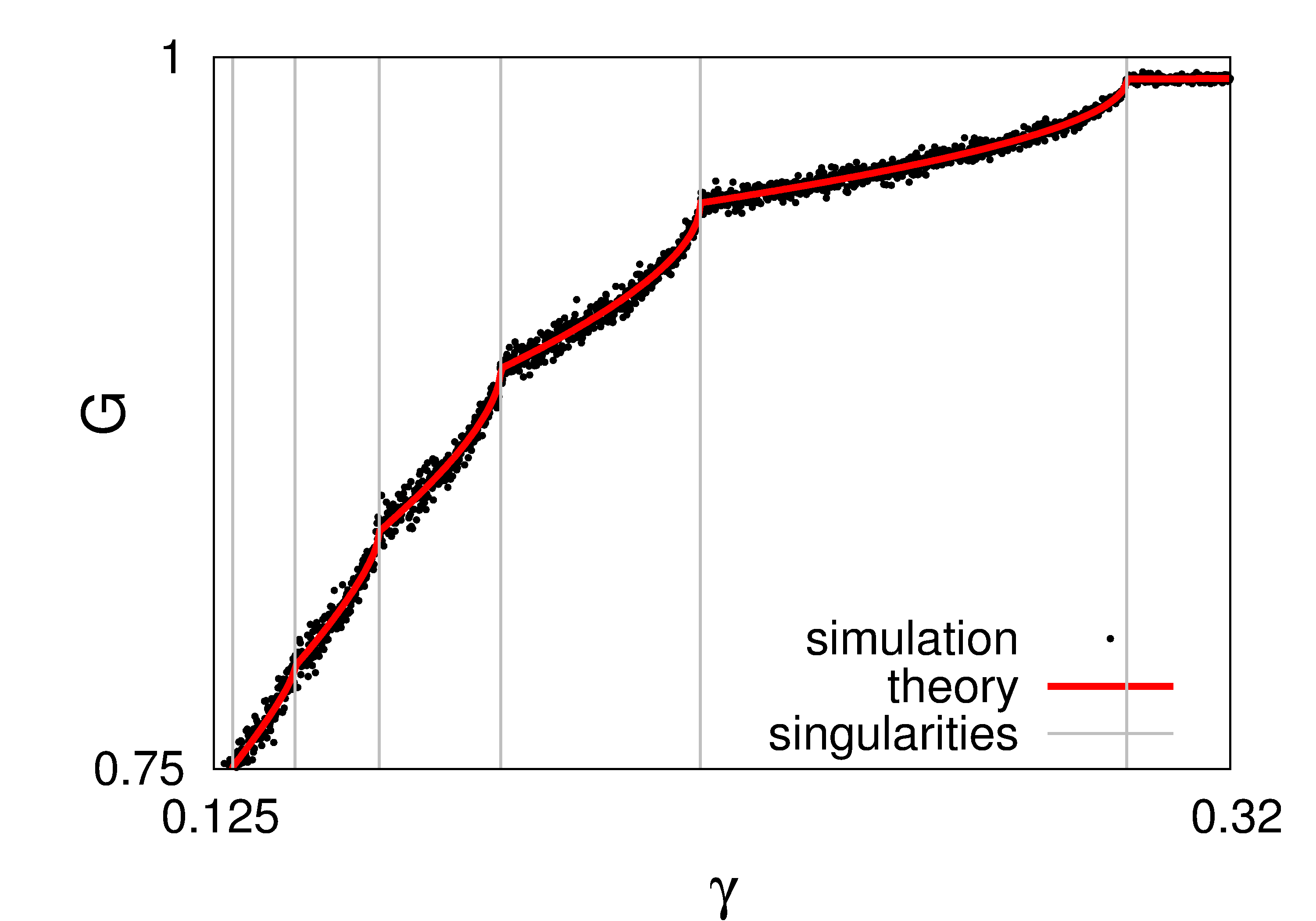}}\subfloat[]{\includegraphics[width=0.45\textwidth]{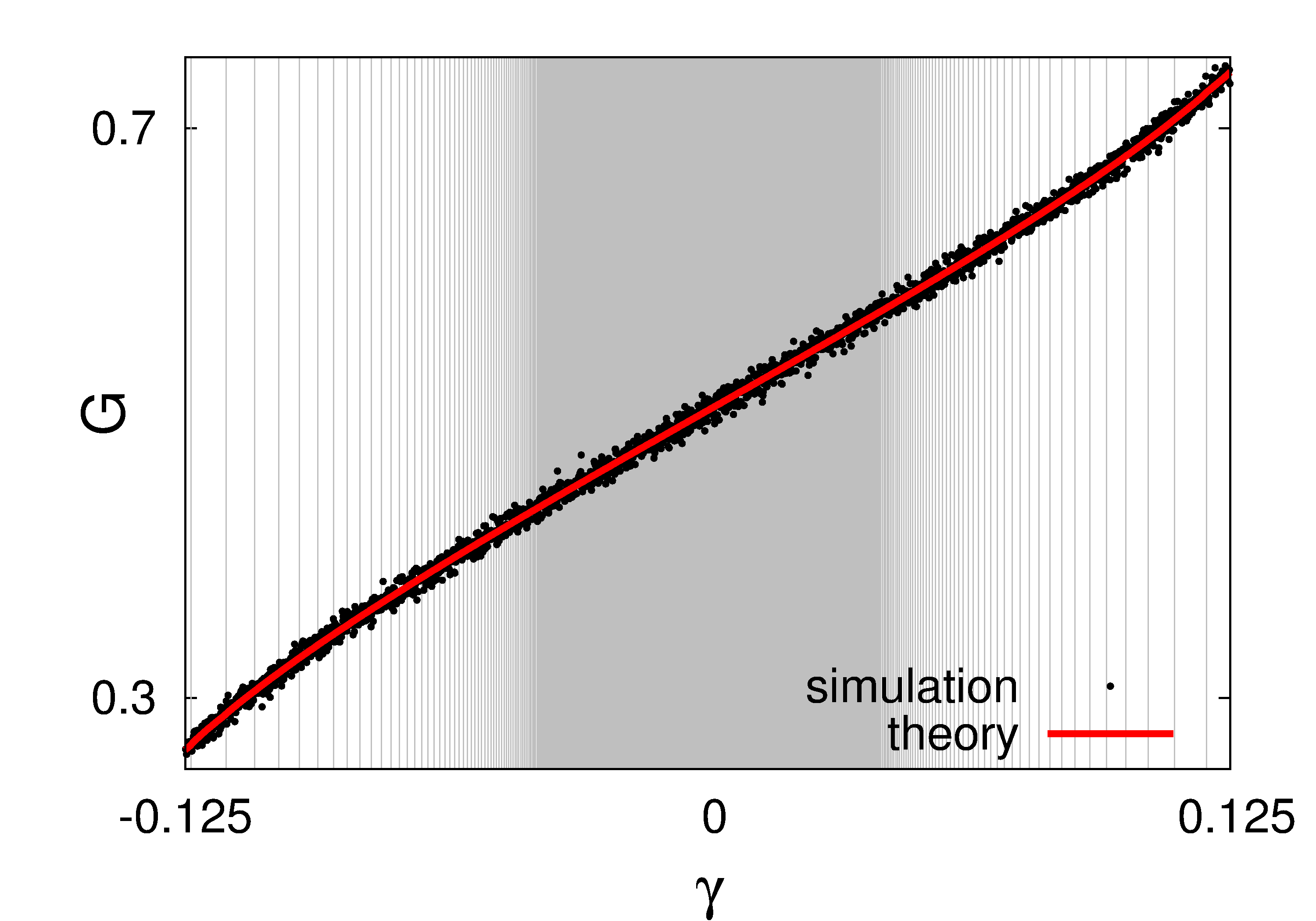}

}
\par\end{centering}

\centering{}\subfloat[]{\includegraphics[width=0.45\textwidth]{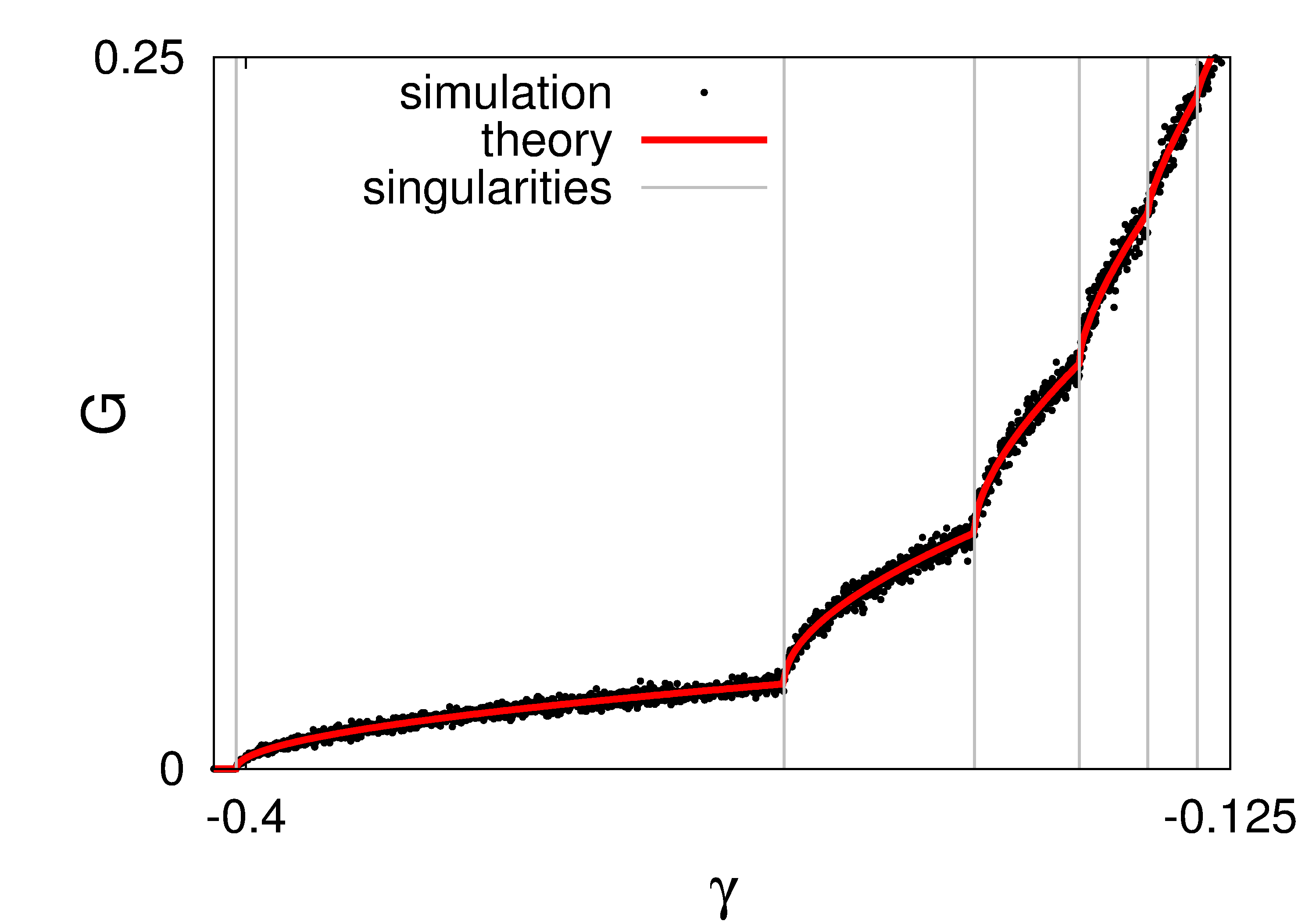}

}\caption{Zooms in Fig.~\ref{fig:Fy}(c).  $G$ is a non-smooth function, showing corners at the same position of the singularities. The corners occur at strong singularities.
More clearly visible corners in the curves $G$ are associated  to more pronounced ``peaks'' in the pdf of $\gamma$.}
\label{fig:Fy-Zoom}
\end{figure}

It is interesting to observe that $G(\gamma)$ is not smooth, i.e., it is not differentiable at the location of the singularities of $g$. 
In particular, the pdf of $\gamma$,
$g(\gamma)$, which is the 
derivative of $G(\gamma)$, has singularities near maxima and
minima of $J_{0}$.  Such singularities are discontinuities
of $g(\gamma)$ and explain the non-smooth character of $G\left(\gamma\right)$,
clearly seen in Figs.~\ref{fig:Fy-Zoom}(a)-(c), that shows zooms in three different regions of Fig.~\ref{fig:Fy}(c). 
The singularities are indicated by vertical lines (in gray color). 
Figures \ref{fig:Fy-Zoom}(a)-(c) show corners  located  at the same position of the singularities. However, not all singularities have clearly visible
corners associated to them.
In Fig.~\ref{fig:Fy-Zoom}(b), though there are many singularities near $\gamma=0$, $G$ is quite smooth near them.
The figures indicate that visible corners appear only in case of strong singularities. As mentioned before, strong singularities are associated to peaks in the pdf of $\gamma$. 
If the derivative of $G$, given by $g$, has a singularity, $G$ has a corresponding corner, but this corner is not necessarily visible. Weak singularities do not produce 
visible corners.

\section{Statistics of confinement}\label{sec:SC}

The study of transport barriers is a fundamental problem in magnetically confined fusion plasmas. 
Since this is a very complex problem involving a wide range of different physical processes, it is  of value to 
study it in simplified settings using reduced models. 
Following this philosophy, in this section  we use the gyro-average standard map to explore the role of finite Larmor radius
effects on transport barriers and escape rates in the presence of drift-waves. 

\subsection{Transport barriers}\label{sec:Pc}

The relative simplicity of the GSM opens the possibility of using results from Hamiltonian dynamical systems to predict the threshold for global transport. In the case when all the particles have the same Larmor radius, $\hat{\rho}^0$, the problem is trivial and reduces to the extensively studied problem of computing the threshold for the destruction of all transport barriers, also knows also as KAM (Kolmogorov-Arnold-Moser) invariant circles, in the standard map with $K=K_0 \hat{\rho}^0$. As it is well known, in this simple case, when $K>K_c=0.9716...$ there are no transport barriers and transport is global \cite{Greene79}. 
In the more realistic case in which the Larmor radii of the particles in the plasma are given by a distribution of the form in Eq.~(\ref{eq:X-PDF}) corresponding to a plasma in thermal equilibrium, the problem is much less trivial because each particle will ``see" a different drift-wave perturbation and as a result it might or might not exhibit global chaos.  In this section we address this problem using the results developed in the previous sections. 
In particular, we compute the probability of global chaos,
i.e. the probability that a given particle will not be confined by  Kolmogorov-Arnold-Moser (KAM) barriers. 

The transition to global chaos occurs when
\begin{equation}
   \left|\gamma\right|>K_{c}/K_{0}
\end{equation}
where, as before, $\gamma = K/K_{0}$, and  $K_{c}=0.9716...$ is the critical parameter that defines the transition to
global chaos in the standard map. As discussed in \cite{Fonseca14}, 
for a given fixed $\hat{\rho}$, increasing $K_{0}$  increases
the effective perturbation parameter $K$  and the amount of chaos in phase space.

Since the sign of $K_{0}$ can be changed by a trivial change in the phase of $\sin \theta_n$, without loss of generality,  we will limit attention to positive values for $K_{0}$. The probability of global chaos, $P_{c}$, is defined as
\begin{equation}
   P_{c}=1-P(-K_{c}/K_{0} \leq \gamma \leq +K_{c}/K_{0})\label{eq:Pc-A}
\end{equation}
where $P(-K_{c}/K_{0}\leq \gamma\leq +K_{c}/K_{0})$ is the probability that  the value of the random variable $\gamma$ is in the interval $-K_{c}/K_{0}\leq \gamma\leq +K_{c}/K_{0}$.  $P_{c}$ provides a measure of  the portion of particles
that can in principle exhibit global chaos. 
Equation (\ref{eq:Pc-A}) can be rewritten as
\begin{equation}
   P_{c}=1-[G(K_{c}/K_{0}) - G(-K_{c}/K_{0})]\label{eq:Pc-B} \, ,
\end{equation}
where $G$ is the cumulative pdf in Eq.~(\ref{eq:Fy_Dist_Def}). 
Substituting (\ref{eq:FyFinalForm-1}) in (\ref{eq:Pc-B}),  and using the fact that $K_{c}/K_{0}>0$,
it is concluded that
\begin{itemize}
   \item If $K_{c}/K_{0} \geq 1$, then $G(K_{c}/K_{0})=1$,  $G(-K_{c}/K_{0})=0$, and
        \begin{equation}
           P_{c} = 0 \, .
            \label{eq:Pc-1} 
        \end{equation}
   \item If $\left|\gamma_{min}\right| \leq K_{c}/K_{0} < 1$, then
         \begin{equation}
            G(K_{c}/K_{0})=
            \sum_{\hat{\rho}^{+}_{i}\in\Gamma_{+K_{c}/K_{0}}}(-1)^{i}\exp\left[-\frac{1}{2}\left(\frac{\hat{\rho}^{+}_{i}}{\hat{\rho}_{th}}\right)^{2}\right],
            \label{eq:gcposit}
         \end{equation}  
         $G(-K_{c}/K_{0})=0$, and
         \begin{equation}
            P_{c} = 1-\exp\left[-\frac{1}{2}\left(\frac{\hat{\rho}^{+}_{0}}{\hat{\rho}_{th}}\right)^{2}\right] \, .
            \label{eq:Pc-2} 
         \end{equation}
         since, in this interval, $\Gamma_{+K_{c}/K_{0}}$ has only one element, $\hat{\rho}^{+}_{0}$.
   \item If $0 < K_{c}/K_{0} < \left|\gamma_{min}\right|$, then Eq. (\ref{eq:gcposit}) also applies for $G(K_{c}/K_{0})$,  $G(-K_{c}/K_{0})$ is given by
         \begin{equation}
            G(-K_{c}/K_{0})=
            \sum_{\hat{\rho}^{-}_{i}\in\Gamma_{-K_{c}/K_{0}}}(-1)^{i}\exp\left[-\frac{1}{2}\left(\frac{\hat{\rho}^{-}_{i}}{\hat{\rho}_{th}}\right)^{2}\right],
            \label{eq:neg}
         \end{equation} 
         and 
         \begin{equation}
            P_{c}=1-\left\{\sum_{\hat{\rho}^{+}_{i}\in\Gamma_{+K_{c}/K_{0}}}(-1)^{i}\exp\left[-\frac{1}{2}\left(\frac{\hat{\rho}^{+}_{i}}{\hat{\rho}_{th}}\right)^{2}\right]-\sum_{\hat{\rho}^{-}_{i}\in\Gamma_{-K_{c}/K_{0}}}(-1)^{i}\exp\left[-\frac{1}{2}\left(\frac{\hat{\rho}^{-}_{i}}{\hat{\rho}_{th}}\right)^{2}\right]\right\}  \, .
            \label{eq:Pc-3}
         \end{equation}
\end{itemize}
Let $S_{K_{c}/K_{0}}$ be the ordered set $( \hat{\rho}^{+}_{0},\hat{\rho}^{-}_{0},\hat{\rho}^{-}_{1},\hat{\rho}^{+}_{1},\hat{\rho}^{+}_{2}, \hat{\rho}^{-}_{2}, \hat{\rho}^{-}_{3}...)$, which corresponds to the set formed by the elements of $\Gamma_{+K_{c}/K_{0}}$ and $\Gamma_{+K_{c}/K_{0}}$. Denoting the terms of
$S$ by $\hat{\rho}_{i}$ such that $\hat{\rho}_{0}=\hat{\rho}^{+}_{0}$, $\hat{\rho}_{1}=\hat{\rho}^{-}_{0}$, $\hat{\rho}_{2}=\hat{\rho}^{-}_{1}$, and so on, Eqs (\ref{eq:Pc-1}), (\ref{eq:Pc-2}) and (\ref{eq:Pc-3}) can be written in the more compact form 
\begin{equation}
   P_{c}=\left\{ \begin{array}{cc}
    0 & 0< K_{0} \leq K_{c}\\ \\
    1-\sum_{\hat{\rho}_{i}\in S_{K_{c}/K_{0}}}(-1)^{i}\exp\left[-\frac{1}{2}\left(\frac{\hat{\rho}_{i}}{\hat{\rho}_{th}}\right)^{2}\right] & 
    K_{c}< K_{0} \, ,
   \end{array}\right.
   \label{eq:Pc_FinalForm}
\end{equation}
where $S_{K_{c}/K_{0}}$ is the set $\{ \hat{\rho}_{i} \}$. of solutions  of $K_{c}/K_{0}=\left|J_{0}(\hat{\rho}_{i})\right|$. 

\begin{figure}[!h]
   \begin{centering}
      \includegraphics[width=0.5\textwidth]{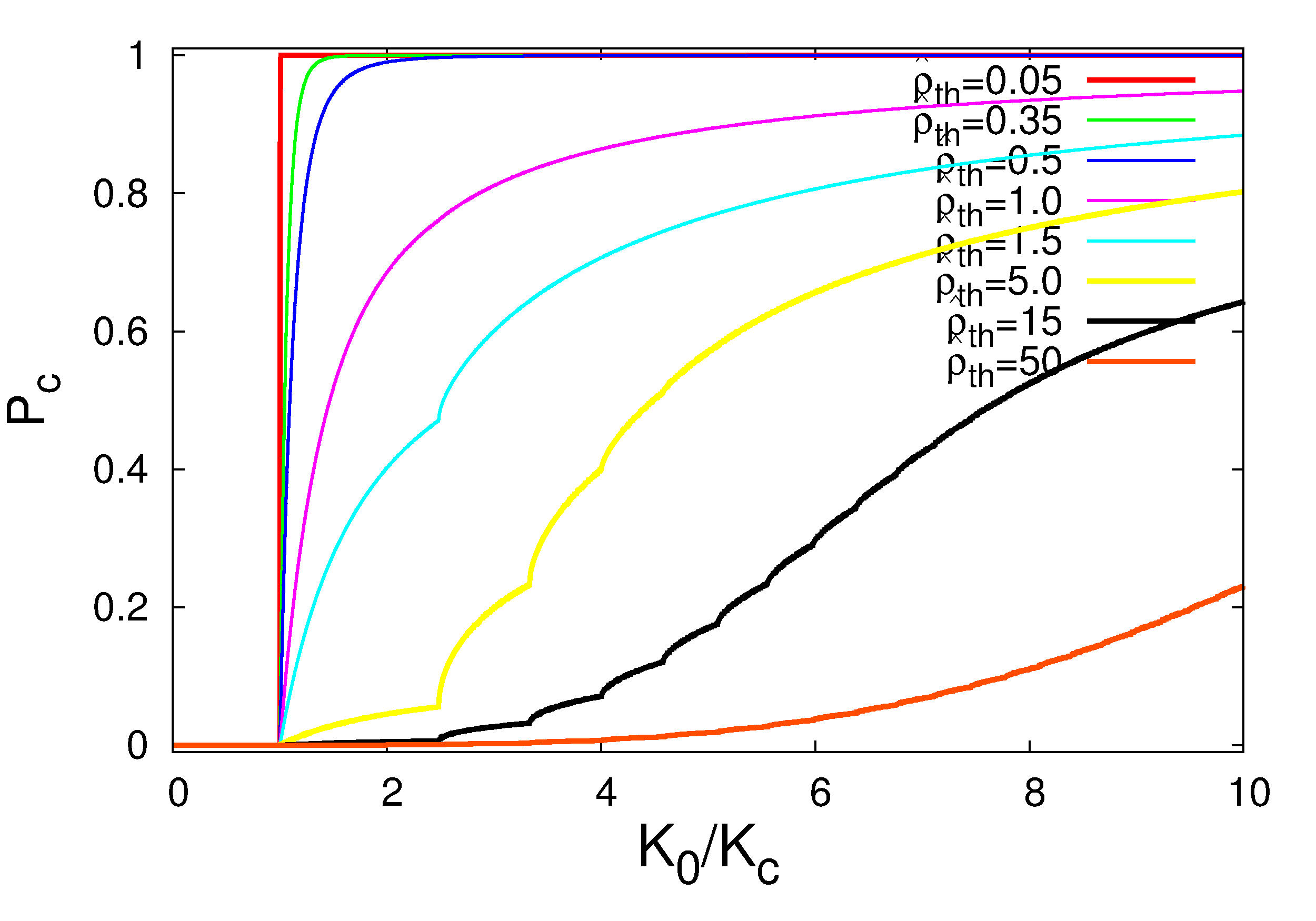}
    \par\end{centering}
    \caption{Probability of global chaos ($P_{c}$)
     as a function of $K_{0}/K_{c}$ and for
     different values of $\hat{\rho}_{th}$. 
     For $0<K_{0}/K_{c}\leq 1$, $P_{c}=0$: particles exhibit quasi-periodic motion,
     are confined inside stability islands, or exhibit chaotic motion bounded by KAM barriers.
     For high and increasing values of $K_{0}/K_{c}$,  $P_{c}$ goes to one, that is, particles following chaotic orbits move freely in
     the radial direction since all KAM barriers are broken.}
     \label{fig:PcIgc}
\end{figure}

Figure \ref{fig:PcIgc} shows $P_{c}$  as a function of $\frac{K_{0}}{K_{c}}$ for
different fixed values of $\hat{\rho}_{th}$.  For $0<K_{0}/K_{c}\leq 1$, $P_{c}=0$ and 
particles can exhibit regular motion (including trapping  inside stability islands and quasi-periodic motion) or chaotic motion. However, in this case all particles remain confined in regions isolated by KAM barriers.
For $K_{0}/K_{c} > 1$, as $K_{0}$ increases,
$P_{c}$ approaches one. That is, most particles move in phase spaces without KAM barriers and those following chaotic orbits move freely in the $I$-direction which corresponds to the radial direction in our simplified drift-wave transport model.
According to Eq. (\ref{eq:Pc_FinalForm}), for large values of $K_{0}/K_{c}$ the sum in (\ref{eq:Pc_FinalForm}) goes to zero. This is because consecutive elements of $S_{K_{c}/K_{0}}$,  $\hat{\rho}_{i}$ and $\hat{\rho}_{i+1}$,
become both near one of the zeros of $\left|J_{0}(\hat{\rho})\right|$. Thus, the exponential terms of the sum, evaluated at $\hat{\rho}_{i}$ and $\hat{\rho}_{i+1}$ and that have opposite signs,
cancel each other.

\begin{figure}
   \begin{centering}
    \includegraphics[width=0.5\textwidth]{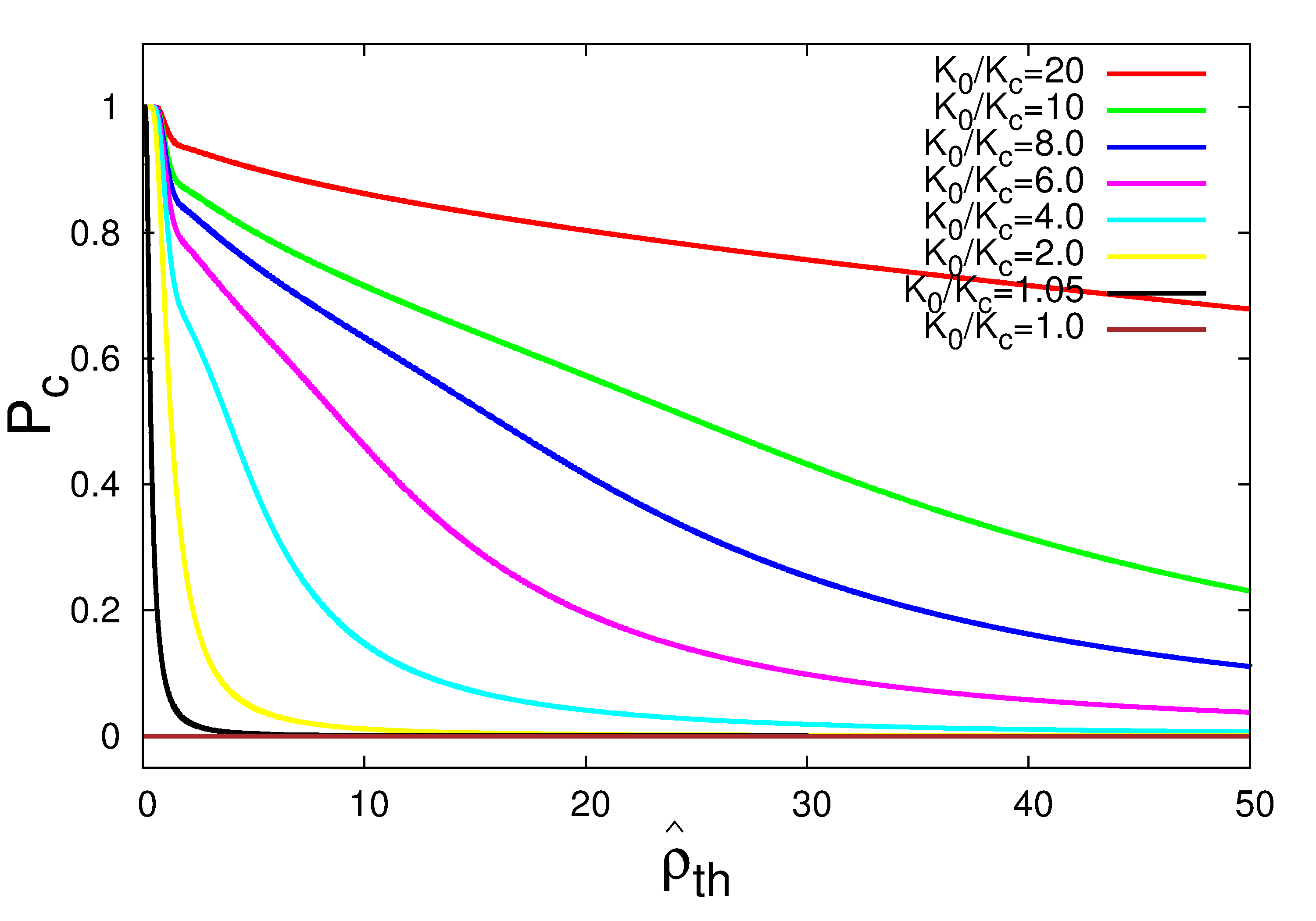}
    \par\end{centering}
    \caption{Probability of global chaos ($P_{c}$) as a function of $\hat{\rho}_{th}$
     and for different fixed values of $K_{0}/K_{c}$.}
    \label{fig:PcRth}
\end{figure}

The probability of global chaos is also plotted
in Fig.~\ref{fig:PcRth} but as a function of $\hat{\rho}_{th}$ for different fixed values of $K_{0}/K_{c}$.
If $0<K_{0}/K_{c}\leq 1$, $P_{c}$ is always zero for any $\hat{\rho}_{th}$,
as indicated by the brown horizontal line ($K_{0}/K_{c} = 1$).
If $K_{0}/K_{c}>1$,  $P_{c}$ is close to one for small $\hat{\rho}_{th}$
and decays for increasing  $\hat{\rho}_{th}$. Still considering the case $K_{0}/K_{c}>1$, the plots shown in Fig.~\ref{fig:PcRth} also indicate a fast decay for small $\hat{\rho}_{th}$
and a slower one for high $\hat{\rho}_{th}$, which, according to Eq. (\ref{eq:Pc_FinalForm}), can   
be explained considering that $dP_{c}/d\hat{\rho}_{th}\sim 1/\hat{\rho}^{3}_{th}$.

\subsection{Escape Rate} \label{sec:escape_rate}

Motivated by the results of subsection \ref{sec:Pc},
in this section we analyze the ``escape rate'', $\eta_{e}$,
 for the standard map and the GSM. In the latter
case, this measure is compared to the probability of global chaos
 The escape rate is computed as follows:
 \begin{itemize}
\item We  construct an ensemble of $N$
particles with initial conditions $(\theta^{i}_0,I_{0})$ where $I_{0}$ is constant and $\{\theta^{i}_0\}_{i=1}^N$ are random numbers uniformly distributed in the interval $(0, 2 \pi)$. 
 
\item For each particle of the ensemble, the map is iterated $T$ times. If for a given $n \leq T$, 
$\left|I_{n}-I_{0}\right|>2\pi$, the iteration stops and the initial condition  $(\theta^{i}_0,I_{0})$ is counted as a escaping orbit.

\item The escaping rate is then computed as $\eta_e= N_{e}/N$ where $N_{e}$ is the total number of escaping orbits in the ensemble. 

\end{itemize}

In the calculations presented here, $I_{0}=\pi$. Although in principle  any other position for the line of initial conditions could be used, some might be more computational efficient than others. 
For example, using $I_{0}$ near $2m \pi$ with $m=0, \pm 1, \pm 2 \ldots$ is not efficient because a significative number of initial conditions  could be located inside the period-one island, which can occupy a relatively large area, even when global chaos is present. In that case, 
it is necessary to wait the maximum number of iterations $T$, making the procedure computationally expensive. 

The escape condition, $\left|I_{n}-I_{0}\right|>2\pi$, is adopted
because the standard map  is invariant under translations by $2\pi$ in the direction of the $I$ coordinate.  This implies that invariant objects like islands, invariant circles, and chaotic orbits repeat themselves under translations by $2\pi$. 
Thus, if a particle initially located at $I_{0}$ is found in a position $I_{n}$ such that the escape condition holds,
 the same orbit will be  by symmetry at $I_{0} \pm 2\pi m$ with $m=1, 2, 3, ...$.

\begin{figure}
\begin{centering}
   \includegraphics[width=0.5\textwidth]{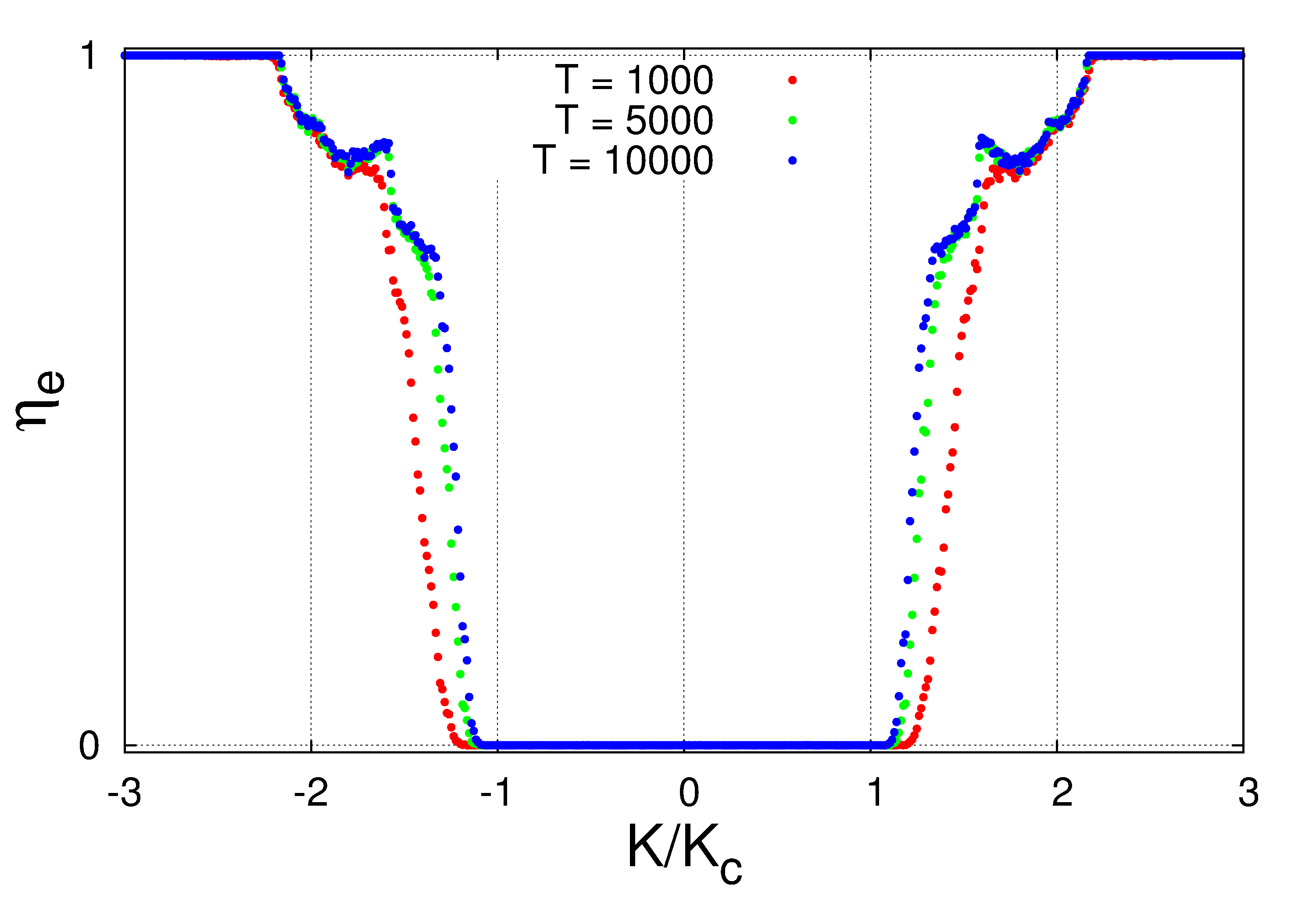}
   \par\end{centering}
   \caption{Rate of escaping particles, $\eta_{e}$, versus $K/K_{c}$ for the standard map model.
   $K$ is the perturbation parameter and $K_{c}$ is the critical parameter
   defining the transition to global chaos.
   If $K > K_{c}$, there are no KAM barriers, allowing particles that follow chaotic orbits to escape. 
   Due to trapping effects around or inside stability islands, there is no abrupt transition in escape rate for 
   $K \simeq K_c$. 
   } 
   \label{fig:ReSmap}
\end{figure}

Figure \ref{fig:ReSmap} shows plots of $\eta_{e}$ versus $K/K_{c}$ in the standard map for different values of the maximum number of iterations $T$, where 
$K$ is the perturbation parameter and $K_{c}=0.971...$ is the critical parameter for
the transition to global chaos. The number of particles used in this calculation was $N=5000$.
 Due to the presence of KAM barriers, 
 no particles can escape if the absolute value of the perturbation is below the critical parameter and thus, as seen in Fig. \ref{fig:ReSmap}, 
 $\eta_{e}=0$ for $|K| \leq K_{c}$.
On the other hand, for $|K|  > K_{c}$ there are no KAM barriers and chaotic orbits  can in principle  escape. However,
due to the trapping of particles inside stability islands and/or long time stickness
of chaotic orbits near islands in practice not all the particles escape and a sharp transition transition from zero to one
at $|K| = K_{c}$ is not observed.
As $|K|/K_c$ increases, the trapping and stickiness is reduced and $\eta_e$ approaches one.

\begin{figure}
   \begin{centering}
      \includegraphics[width=0.53\textwidth]{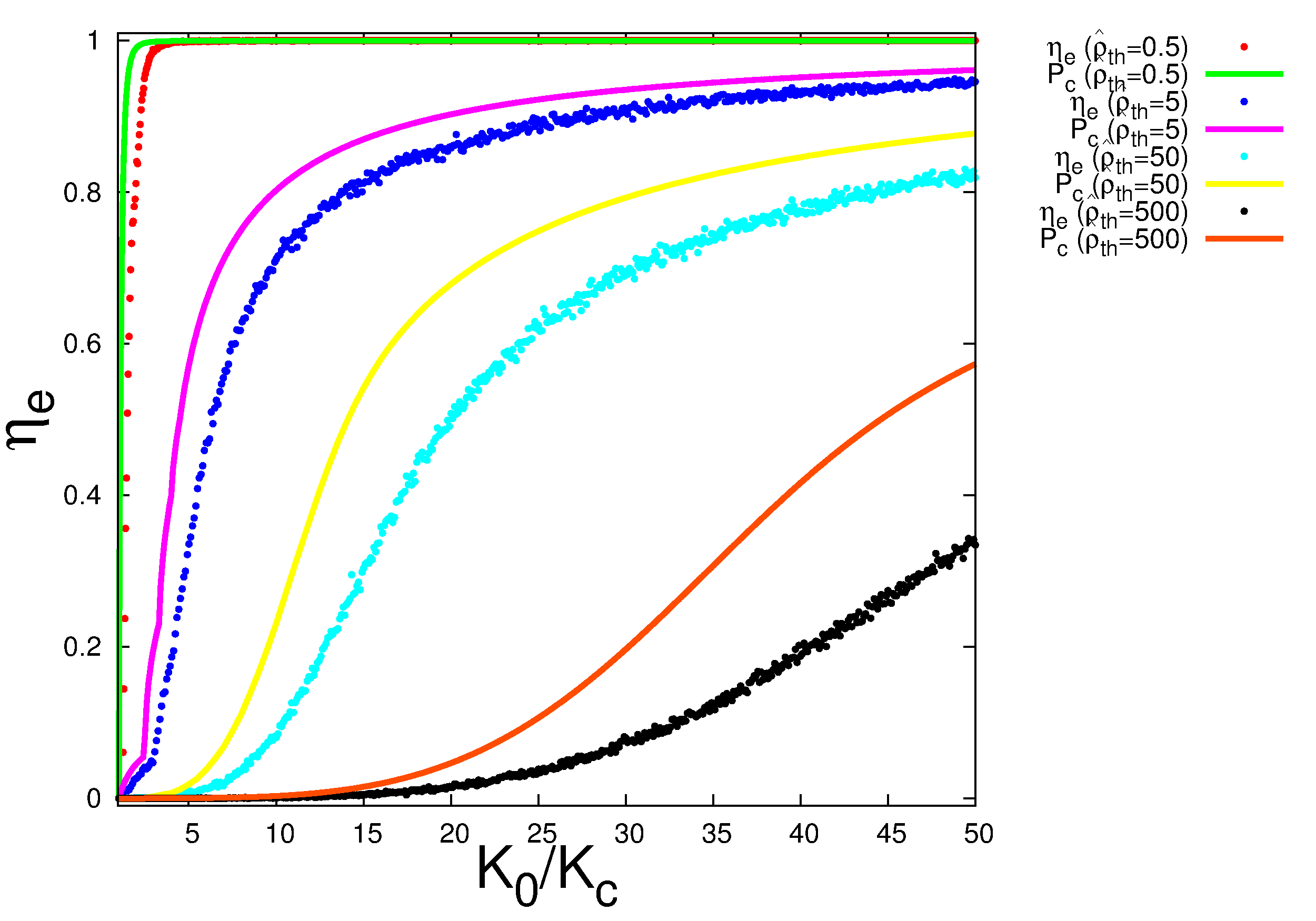}
   \par\end{centering}
   \caption{GSM model's escape rate (colored points) for increasing
    $K_{0}/K_{c}$ and fixed $\hat{\rho}_{th}$  compared to the probability of
    global chaos (colored continuous lines). The analytical quantity $P_{c}$
    provides an upper bound for $\eta_{e}$. Parameters: $N=5\times10^{3}$; $T=5\times10^{3}$.}
    \label{fig:ReVarKo}
\end{figure}

Figure \ref{fig:ReVarKo} shows the escape rate, $\eta_e$, as function of $K_{0}/K_{c}$ for different values of $\hat{\rho}_{th}$ in the GSM with the Maxwellian distribution of gyro-radius in Eq.~(\ref{eq:X-PDF}).   
Each point corresponds to a simulation where $\eta_{e}$ is calculated for an ensemble of $N=5\times10^{3}$ particles and a maximum number of iterations $T=5\times10^{3}$. The figure also shows
the probability of global chaos (colored continuous lines) according to the analytical formula in Eq.~(\ref{eq:Pc_FinalForm}) 
for the same  values of $\hat{\rho}_{th}$. The results show that the analytical quantity $P_{c}$
provides an upper bound for $\eta_{e}$.  
This is consistent with the fact that $P_{c}$ quantifies the probability that  a particle can escape in {\em principle}
whereas $\eta_e$ quantifies the probability that  a particle escapes in {\em practice}. As mentioned before, even when a particle could escape because the effective perturbation parameter is large enough, the particle might not escape if it is trapped inside an stability island. 
However, the difference diminishes with increasing $K_{0}$. According to 
Eq. (\ref{eq:Kaverage}), the mean effective perturbation $\left\langle K\right\rangle$ increases proportionally with $K_{0}$, 
suppressing islands' effects and increasing chaos.

\begin{figure}
   \begin{centering}
     \includegraphics[width=0.45\textwidth]{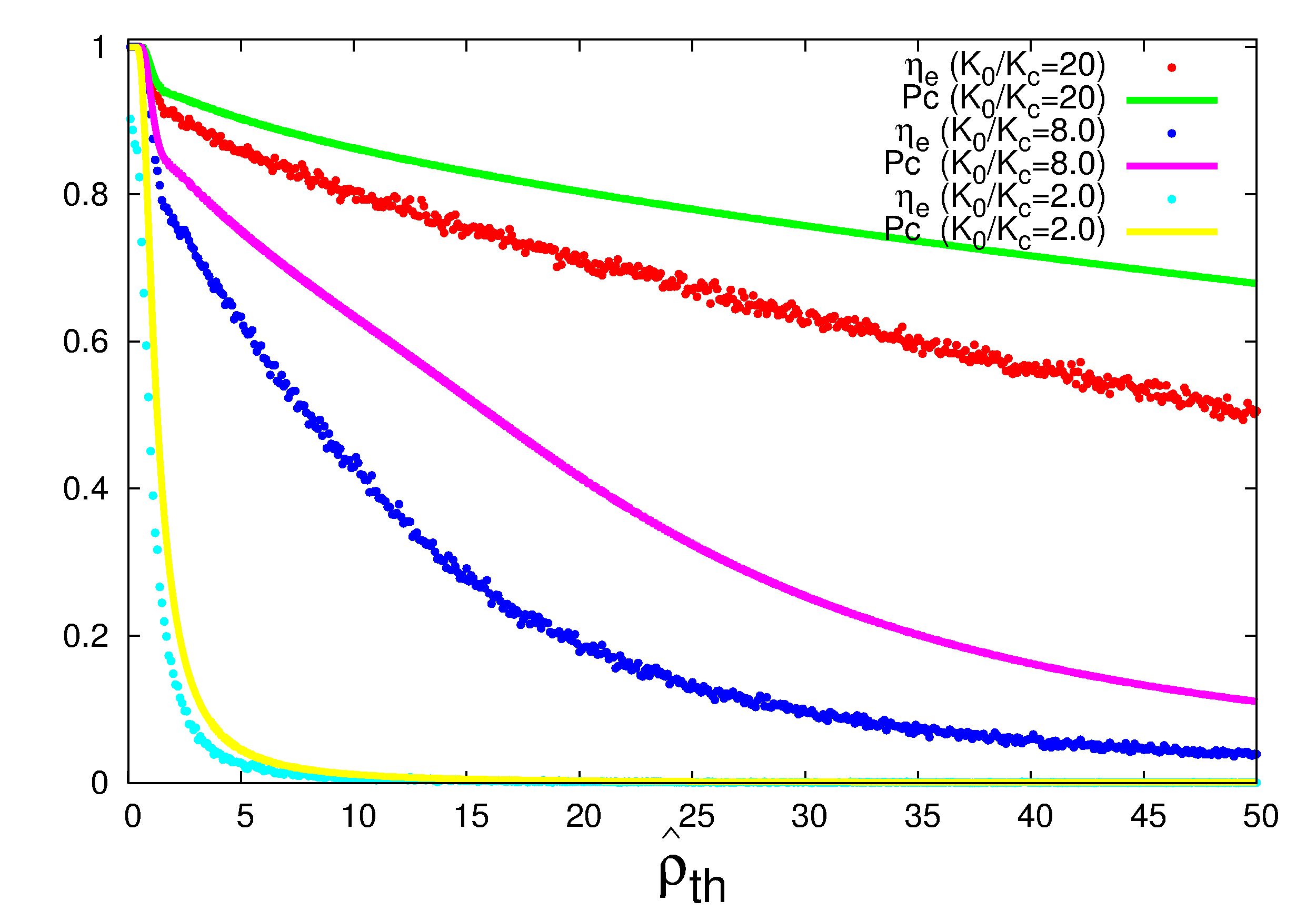}
   \par\end{centering}
   \caption{Rate of escaping particles and probability of global chaos in the GSM model for increasing
   $\hat{\rho}_{th}$ and fixed $K_{0}/K_{c}$. Parameters: $N=5\times10^{3}$; $T=5\times10^{3}$.}
   \label{fig:ReTh}
\end{figure}

Figure \ref{fig:ReTh} shows plots of the GSM model's escape rate and probability of global chaos
for varying $\hat{\rho}_{th}$ and fixed $K_{0}/K_{c}$. Again, the escape rate is below the limit given by the probability of global chaos. 
It is expected that both measures go to zero with increasing $\hat{\rho}_{th}$.
As prescribed by Eqs. (\ref{eq:Kaverage}) and (\ref{eq:Kdispersion}), the average and dispersion of the effective perturbation
go to zero for high values of $\hat{\rho}_{th}$, restoring islands and KAM barriers. This
is reflected, for example, in the fast decaying plots of Fig. \ref{fig:ReTh} ($K_{0}/K_{c}=2.0$).   

\section{Statistics of particle trapping}\label{sec:Rt}

Particle trapping is an ubiquitous phenomena in transport driven by plasma waves. In the co-moving reference of a traveling wave, trapping results from the confinement of particles at the minimum of the potential well. In the  simple standard map description this corresponds to the period-one island with elliptic fixed point located at $I=0$ and $\theta=\pi$. 

In the GSM model this problem is more complex  because, as explained before, the effective drift-wave amplitude, 
and as result the stability of the period-one island fixed point, depend on the statistics of the  Larmor radii.  
As a result, depending on their Larmor radius, 
some particles ``see'' phase spaces
where the fixed points are hyperbolic, and others where the fixed points are parabolic or elliptic.  In general,
particles located near elliptic fixed points are trapped by the corresponding period-one islands,
and those near hyperbolic points spread in their respective phase spaces.  

In this section, we study this problem by studying the \emph{rate of trapping}, $\eta_{t}$, computed as follows:
\begin{itemize}

\item  We  construct an ensemble of $N$
particles with random initial conditions uniformly distributed on a disk of radius $\epsilon$ centered at the 
location of the O-point, $(I,\theta)=(0,\pi)$.

\item For each particle of the ensemble, the map is iterated $T$ times. If for a given $n \leq T$,
the particle escapes from a concentric circle of radius $r \gg \epsilon$  the iteration stops and the initial condition   is counted as not trapped.

\item The trapping rate is then computed as $\eta_t= N_{t}/N$ where $N_{t}$ is the total number of trapped orbits, i.e. the total number of particles that remained in the disk of radius $r$ after $T$ iterations.  
\end{itemize}

Note that, because of the translation invariance of the map, the method can be applied to other fixed points
located at $\theta=\pi$ and $I= \pm 2\pi m$, where $m=1, 2, 3, ...$ .  
In all simulations presented here, we used $\epsilon=0.05$, $r=1.0$,  $N=5\times10^{3}$, and 
$T=5\times10^{3}$. Different values can also be used under the condition of keeping $\epsilon \ll r$, $r \leq \pi$ and
using large values for $N$ and $T$.
The condition $r \leq \pi$ comes from the fact that
the standard map is modulated by $2\pi$ in the direction of coordinate $\theta$.

\begin{figure}
\begin{centering}
\includegraphics[width=0.5\textwidth]{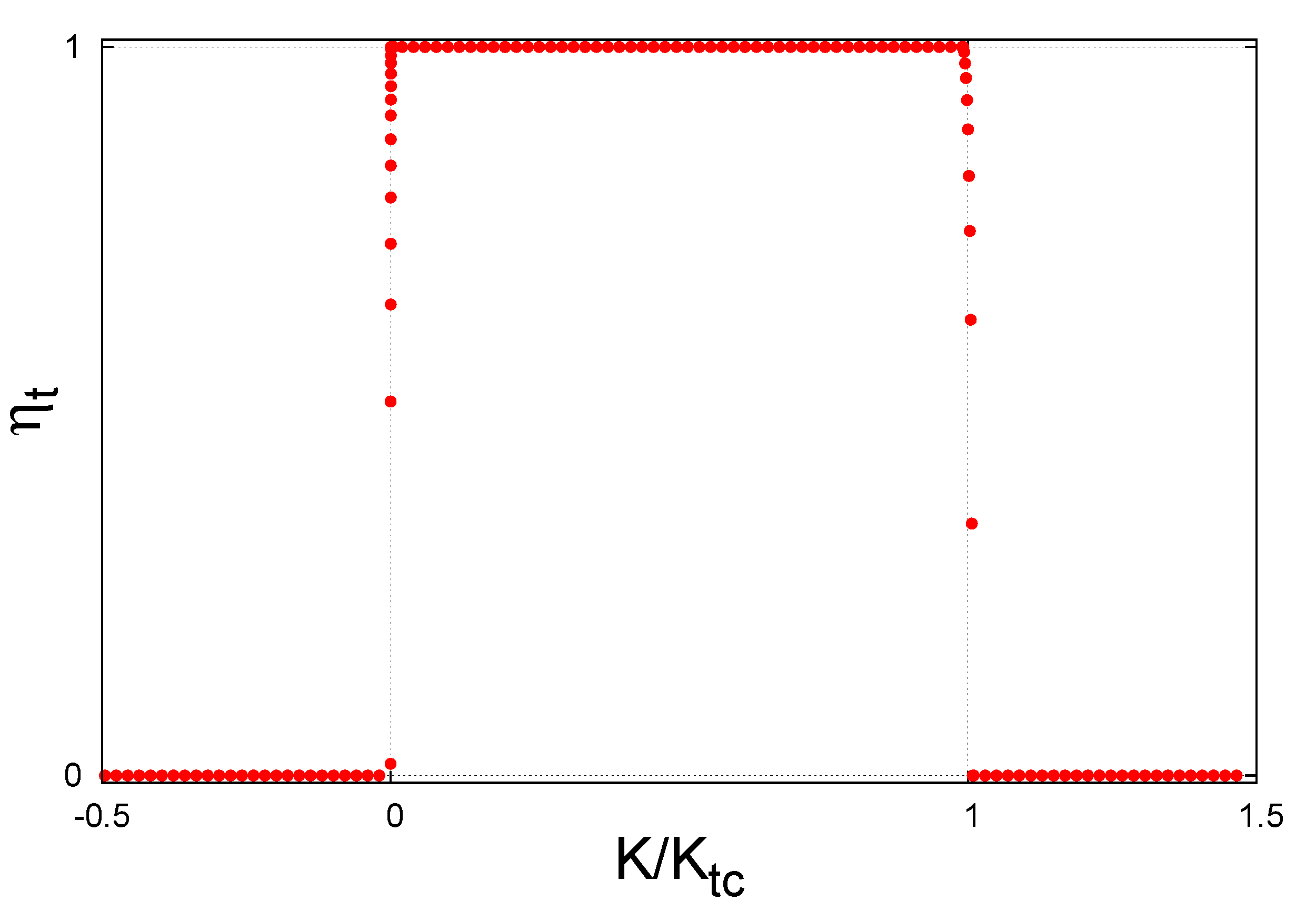}
\par\end{centering}

\caption{Rate of trapping in the standard map by the period-one island associated to the
fixed point located at $\theta_{0}=\pi$ and $I_{0}=0$. Two abrubt transitions occur near $0$ and $K_{tc}\simeq 4.04$.
The trapping interval $0<K<K_{tc}$ corresponds approximately to the stability interval of 
the fixed point, which is elliptic for $0<K<4$ and hyperbolic for $K<0$ and $K>4$.}
\label{fig:RtIgcSmap}
\end{figure}

Figure \ref{fig:RtIgcSmap} shows the trapping rate in the standard map (red points) versus $K/4$, where $K$
denotes the perturbation parameter. 
Two abrupt transitions are observed near $0$ and $K \simeq 4.04$. 
For $0<K<4.04$, orbits remain  trapped ($\eta_{t}=1$) during $T=5\times10^{3}$ iterations,
whereas for $K<0$ or $K>4.04 $ all of them
escape ($\eta_{t}=1$). This result is consistent with the well-known result that the O-point of the period-one island in the standard map losses stability at $K_{tc}=4$ and the period-one fixed point is unstable for $K<0$. 

Before analyzing the rate of trapping in the GSM model, 
we define the \emph{probability of trapping}, $P_{t}$, as the probability of a particle 
to have an effective perturbation parameter $K$ in the trapping interval $0<K<K_{tc}$. That is 
\begin{equation}
   P_{t} = P(0<\gamma<K_{tc}/K_{0}),\label{eq:Pt-A}
\end{equation}
where $K_{0}>0$ and  $P$ denotes the probability that  the value of the random variable $\gamma$ is in the interval 
  $0<\gamma<K_{tc}/K_{0}$. In term of the cumulative distribution function $G$ in Eq.~(\ref{eq:Fy_Dist_Def}), 
\begin{equation}
   P_{t} = G(K_{tc}/K_{0})-G(0).\label{eq:Pt-B}
\end{equation}
Using  Eq.(\ref{eq:FyFinalForm-1}), we have
\begin{equation}
   \label{eq:Pt-Eq1}
   P_{t} = 1-\sum_{\hat{\rho}_{j}\in\Gamma_0}(-1)^{i}\exp\left[-\frac{1}{2}\left(\frac{\hat{\rho}_{j}}{\hat{\rho}_{th}}\right)^{2}\right], \quad  {\rm for} \quad 0 < K_{0}/K_{tc}\leq1
\end{equation}
and
\begin{equation}
   P_{t} = \sum_{\hat{\rho}_{i}\in\Gamma_{K_{tc}/K_{0}}}(-1)^{i}\exp\left[-\frac{1}{2}\left(\frac{\hat{\rho}_{i}}{\hat{\rho}_{th}}\right)^{2}\right]-\sum_{\hat{\rho}_{j}\in\Gamma_0}(-1)^{j}\exp\left[-\frac{1}{2}\left(\frac{\hat{\rho}_{j}}{\hat{\rho}_{th}}\right)^{2}\right], \quad {\rm for} \quad K_{0}/K_{tc}>1 
   \label{eq:Pt-Eq2}
\end{equation}

\begin{figure}
   \begin{centering}
      \includegraphics[width=0.5\textwidth]{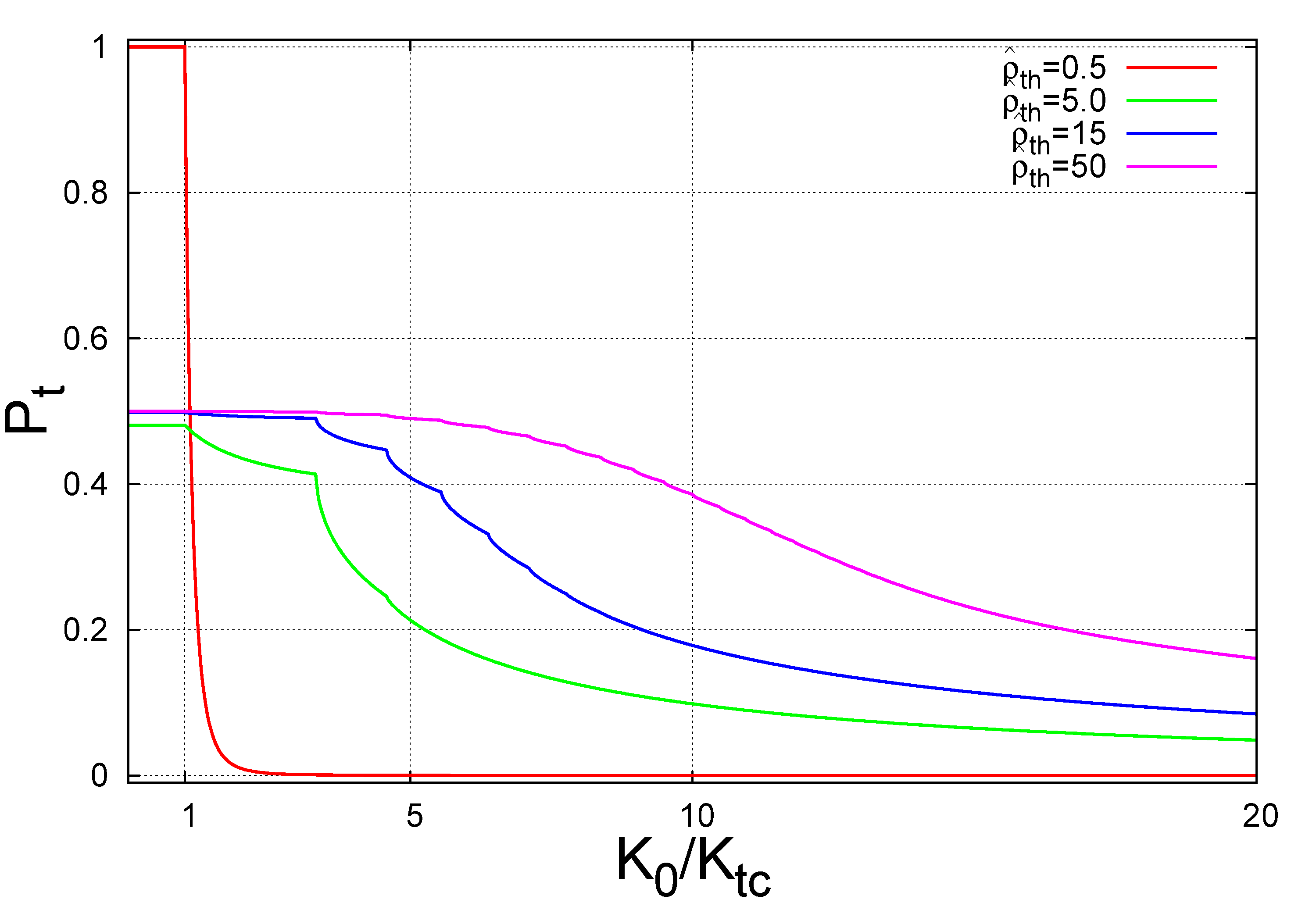}
    \par\end{centering}
    \caption{Probability of trapping, $P_{t}$,  versus $K_{0}/K_{c}$.
     $P_{t}$ is constant for $K_{0}/K_{tc}<1$ and goes to zero for $K_{0}/K_{tc} \gg 1$.} 
    \label{fig:PtIgc}
\end{figure}

Figure \ref{fig:PtIgc} shows plots of the probability of trapping versus $K_{0}/K_{tc}$ for different values  of $\hat{\rho}_{th}$.
In agreement with the fact that
Eq. (\ref{eq:Pt-Eq1}) has no dependence on $K_{0}$,  $P_{t}$ is constant for $K_{0}/K_{tc}<1$.
If $K_{0}/K_{tc}>1$, $P_{t}$ decreases with increasing $K_{0}$.
According to Eq. (\ref{eq:Pt-Eq2}), when $K_{0}/K_{tc} \gg 1$, $\Gamma_{K_{c}/K_{0}} \rightarrow \Gamma_0$, i.e. the two sets
become nearly equivalent and the corresponding sums cancel each other.
Thus, $P_{t}$ goes to zero for high values of $K_{0}/K_{tc}$.

\begin{figure}
   \begin{centering}
      \includegraphics[width=0.5\textwidth]{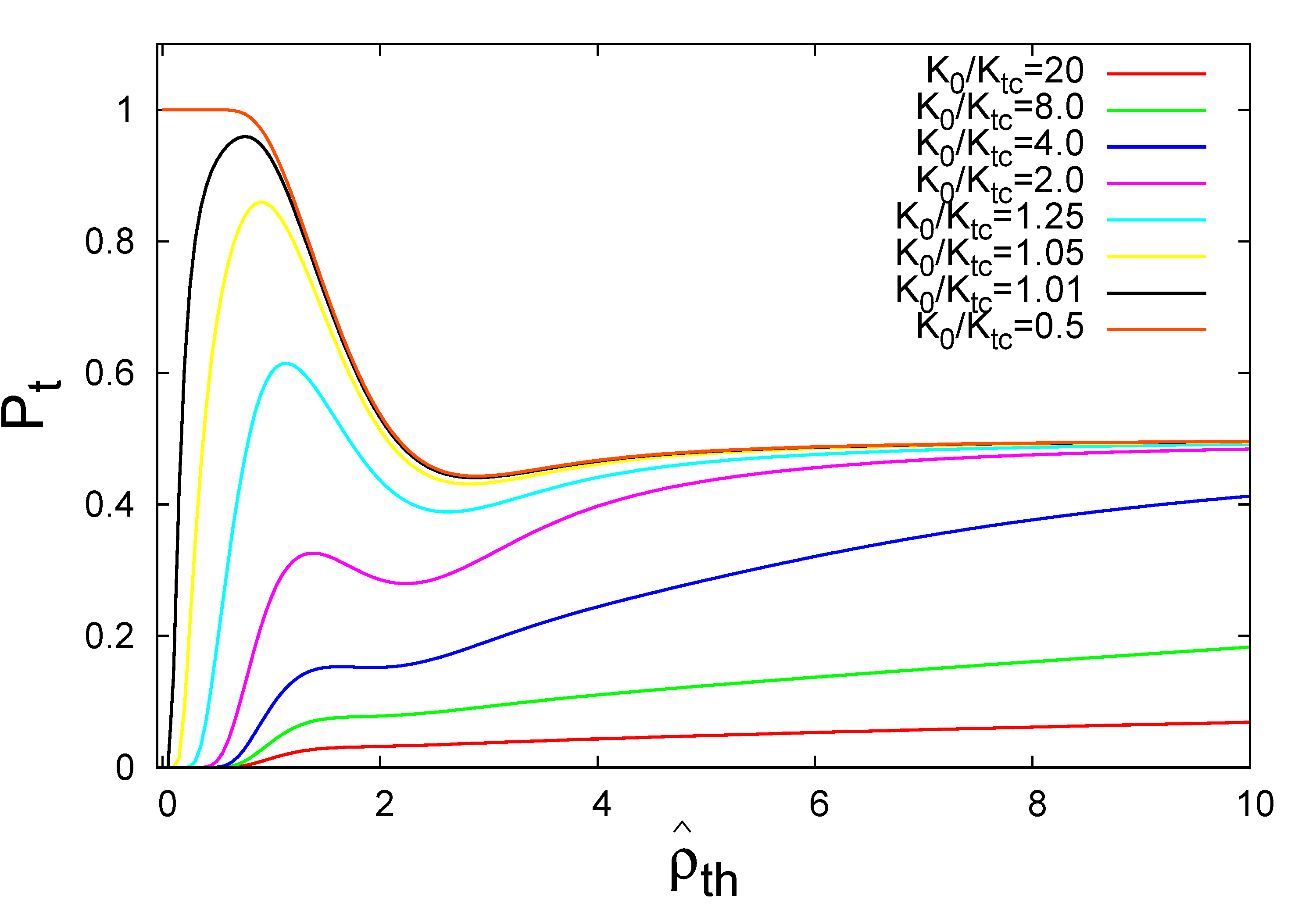}
   \par\end{centering}
   \caption{Probability of trapping ($P_{t}$) versus $\hat{\rho}_{th}$ for different values of $K_{0}/K_{tc}$.}
   \label{fig:PtRth}
\end{figure}

Figure \ref{fig:PtRth} shows plots of the probability of trapping versus $\hat{\rho}_{th}$ for different values of $K_{0}/K_{tc}$.
If $K_{0}/K_{tc}=0.5$,  $P_{t}\simeq1$ for small values of $\hat{\rho}_{th}$, decreases to a minimum and 
increases to a constant level near $0.5$. 
If $K_{0}/K_{tc}>1$, the plots exhibit the following features:
$P_{t}$ increases from zero to a maximum, decreases to a local minimum and increases 
until reaching again the same level near $0.5$. The limit case of small $\hat{\rho}_{th}$ values can be understood through the fast exponential decaying
terms of Eqs. (\ref{eq:Pt-Eq1}) and (\ref{eq:Pt-Eq2}). If $\hat{\rho}_{th} \rightarrow 0$, these terms go to zero and then
we have: $P_{t} \rightarrow 1$ for $0<K_{0}/K_{tc} \leq 1$ and $P_{t} \rightarrow 0$ for $K_{0}/K_{tc}>1$.
As  mentioned before, the plots in Fig. \ref{fig:PtRth} show that $P_{t}$ becomes constant 
at high $\hat{\rho}_{th}$. This is iconsistent with Eqs. (\ref{eq:Pt-Eq1}) and (\ref{eq:Pt-Eq2})
since their derivatives with respect to $\hat{\rho}_{th}$ vary  like $dP_{t}/d\hat{\rho}_{th}\sim 1/\hat{\rho}^{3}_{th}$ 
and approach zero for large $\hat{\rho}_{th}$.

\begin{figure}
   \begin{centering}
      \includegraphics[width=0.5\textwidth]{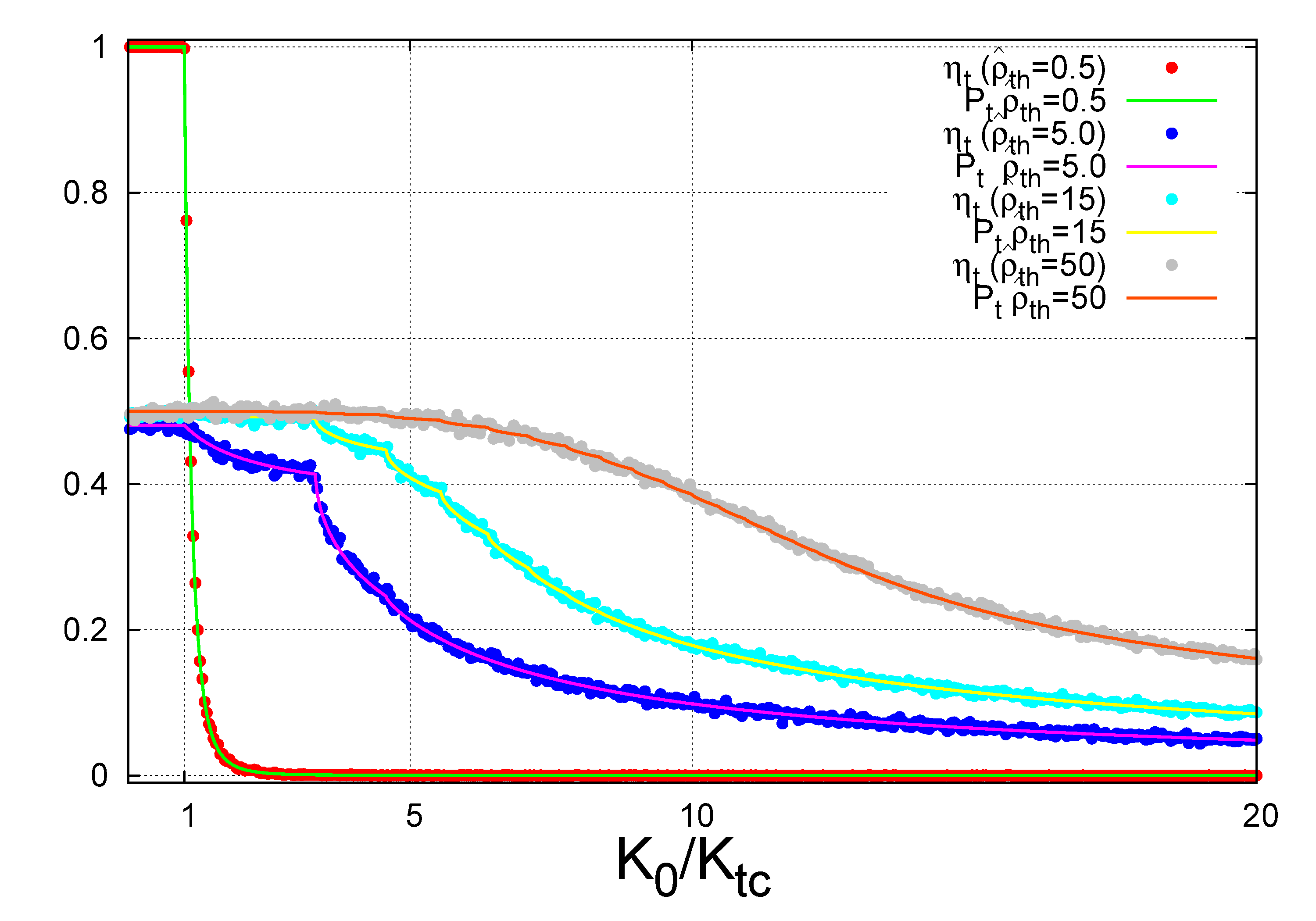}
   \end{centering}
   \caption{GSM's rate of trapping (points) versus $K_{0}/K_{tc}$ for fixed values of $\hat{\rho}_{th}$.
    The probability of trapping (lines), defined by Eqs. (\ref{eq:Pt-Eq1}) and (\ref{eq:Pt-Eq2}), provides a 
    good analytical estimate for $\eta_{t}$.
    Parameters: $N=5\times10^{3}$; $T=5\times10^{3}$; $\epsilon=0.05$; $r=1$. }
   \label{fig:RtIgc}
\end{figure}

\begin{figure}
   \begin{centering}
      \includegraphics[width=0.5\textwidth]{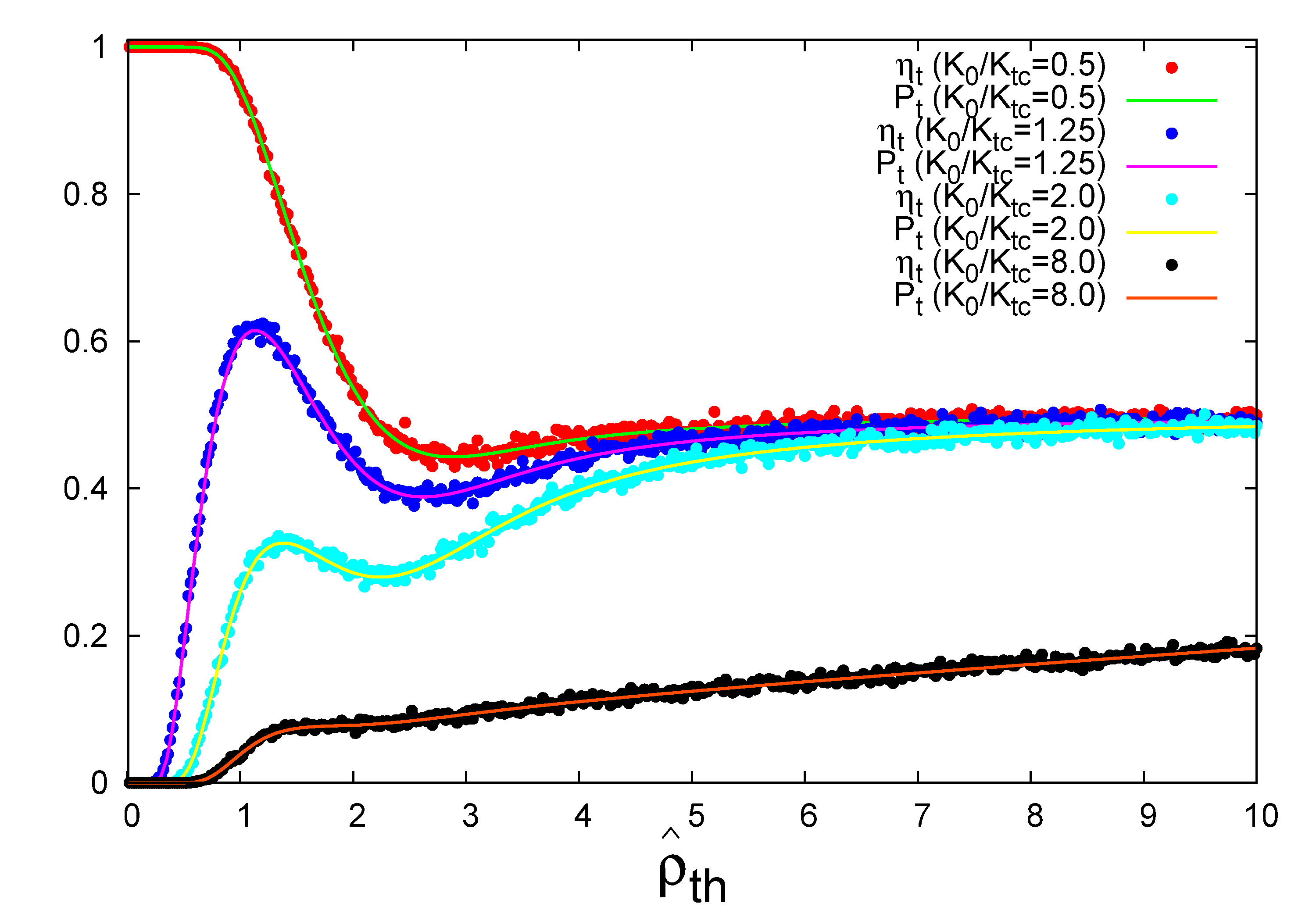}
   \end{centering}
   \caption{$\eta_{t}$ (points)  versus $\hat{\rho}_{th}$ for different fixed values of $K_{0}/K_{tc}$.  $\eta_{t}$ is compared to $P_{t}$
   (lines). Both results are in good agreement.
    Parameters: $N=5\times10^{3}$; $T=5\times10^{3}$; $\epsilon=0.05$; $r=1$.}
   \label{fig:RtRth}
\end{figure}

Figures \ref{fig:RtIgc} and \ref{fig:RtRth} show the 
numeric results (colored points) of the rate of trapping in the GSM model
 computed for an ensemble of $N=5\times10^{3}$ particles with random Larmor radii distributed according to Eq.~(\ref{eq:X-PDF}).  $N$ gyro-averaged standard maps, with different effective perturbation parameters
and the same $K_{0}$, are iterated up to $T=5\times10^{3}$ times.
The value of $K_{0}$ is defined by the product between a given ratio $K_{0}/K_{tc}$, whose values
are shown in the horizontal axis, and the critical parameter $K_{tc}$, estimated as $K_{tc}\simeq 4.04$. 
The initial positions of the particles are randomly located near point $O$ inside 
a circular region of radius $\epsilon$. Again, we adopt $\epsilon=0.05$, and the trapping circular region has a radius defined by $r=1$.
The plots of the rate of trapping in the GSM are compared to
plots of the probability of trapping, also shown in Figs. \ref{fig:PtIgc} and \ref{fig:PtRth}.
Very good agreement is observed between the analytical and the numerical results.

Some properties of the rate of trapping shown in Fig. \ref{fig:RtIgc} can be understood by
analyzing the average of the effective perturbation, $\left\langle K\right\rangle$.
According to Eq. (\ref{eq:Kaverage}),
for any $\hat{\rho}_{th}>0$, $\left\langle K\right\rangle/K_{0} \leq 1$ and,
if $\hat{\rho}_{th}$ is kept fixed, $\left\langle K\right\rangle = O(K_{0})$.
Thus,  $0 < K_{0} < K_{tc}$ implies that $0 < \left\langle K\right\rangle < K_{tc}$, and particles, in average, 
are trapped by the period-one island. 
This explains why, even varying $K_{0}$ inside the trapping interval $0 < K_{0} < K_{tc}$, the rate of
trapping remains approximately constant.      
However, increasing $K_{0}$ indefinitely moves the average of the effective perturbation outside the trapping interval in order
that the rate of trapping starts to decay if $K_{0}>K_{tc}$, as can also be seen in Fig. \ref{fig:RtIgc}.  
 
Figure \ref{fig:RtRth} shows that, for small $\hat{\rho}_{th}$, the rate of trapping  exhibits different behavior for $0<K_{0}<K_{tc}$ and  $K_{0}>K_{tc}$. 
Consider first the case $K_{0}=0.5K_{tc}$ 
(red points). According to Eqs. (\ref{eq:Kaverage}) and (\ref{eq:Kdispersion}), 
if $\hat{\rho}_{th} \rightarrow 0$, then $\left\langle K\right\rangle \rightarrow K_{0}$,
$\sigma_{K}^{2} \rightarrow 0$ and, since $0<K_{0}<K_{tc}$,  most particles are trapped. 
This property is observed in Fig. \ref{fig:RtRth}, where  $\eta_{t} \simeq 1$ for small
$\hat{\rho}_{th}$ and $K_{0}=0.5K_{tc}$ .
On the other hand, when $K_{0}>K_{tc}$, we have again $\left\langle K\right\rangle \rightarrow  K_{0}$ and
$\sigma_{K}^{2} \rightarrow  0$ for $\hat{\rho}_{th} \rightarrow 0$, but most $K$ values are outside the trapping interval. Thus,
$\eta_{t} \simeq 0$ for small $\hat{\rho}_{th}$ and $K_{0} > K_{tc}$ .
However, since $\left\langle K\right\rangle \sim \exp\left(-\hat{\rho}_{th}^{2}/2\right)$, increasing $\hat{\rho}_{th}$ moves
$\left\langle K\right\rangle$ inside the trapping interval, increasing then the number of trapped particles.

In most cases shown in Fig. \ref{fig:RtRth}, we observe the occurrence of local minima in the rate of trapping.    
This property can be explained by the variation of the dispersion of the effective perturbation with increasing
$\hat{\rho}_{th}$. We've seen that, according to Eq. (\ref{eq:Kdispersion}), $\left\langle K\right\rangle$, keeping $K_{0}$ fixed,
varies in the same way as $\sigma_{\gamma}^{2}$, increasing from zero until reaching a maximum and then decreasing to zero again.
If $\left\langle K\right\rangle$ is inside the trapping
interval and the dispersion increases, then the number 
of values of the effective perturbation outside the trapping interval also increases. If the dispersion reaches its maximum,
then the rate of trapping reaches a minimum. 
Since $\left\langle K\right\rangle$ remains inside the trapping interval
($\left\langle K\right\rangle \rightarrow 0^{+}$ for $\hat{\rho}_{th} \rightarrow +\infty$), 
if the dispersion starts to decrease from its maximum, then values of the effective perturbation are brought back to the trapping interval
and then the rate of trapping starts to increase. 

A final comment about Fig. \ref{fig:RtRth} refers to the constant level $\eta_{t} \simeq 0.5$, reached at high values of 
$\hat{\rho}_{th}$. Let $I_{\epsilon}$ be a small neighborhood of zero, 
defined by $I_{\epsilon}=(-\epsilon,+\epsilon)$ and such that  $\epsilon$ is positive and arbitrarily small.
The probability that $\gamma$ is positive given that $\gamma \in I_{\epsilon}$ is the conditional probability $P_{+}=P(0<\gamma<+\epsilon |-\epsilon < \gamma < +\epsilon)$, which can be written as: 
\begin{equation}
   P_{+} = \frac{G(\epsilon)-G(0)}{G(\epsilon)-G(-\epsilon)}. 
   \label{eq:Pplus} 
\end{equation}
Since $G(\pm \epsilon)=G(0) \pm \epsilon g(0) + O(\epsilon^2)$, we see that $P_{+}=\epsilon g(0)/2\epsilon g(0)=1/2$. 
Therefore, if  $\gamma$ has values near zero, these values occur with equal probabilities inside 
$-\epsilon < \gamma < 0$ and $0<\gamma<+\epsilon$.
For $\hat{\rho}_{th} \rightarrow +\infty$, $\left\langle \gamma\right\rangle \rightarrow 0$, $\sigma_{\gamma}^{2} \rightarrow 0$,
resulting that values of $\gamma$ become concentrated in a small neighborhood $I_{\epsilon}$, half of them in $(-\epsilon,0)$ and the other half in $(0,+\epsilon)$. 
Thus, increasing $\hat{\rho}_{th}$ makes values of the effective perturbation $K$ to concentrate inside 
$(-\epsilon K_{0},+\epsilon K_{0})$, half outside the trapping interval ($-\epsilon K_{0} < K < 0$) and the other half inside the 
trapping interval ($0<K<\epsilon K_{0} << K_{tc}$). This means that  $\eta_{t}$ goes to $1/2$ 
for high $\hat{\rho}_{th}$, as shown in Fig. \ref{fig:RtRth}.

\section{Summary and Conclusions} \label{sec:conclusion}

We have presented a statistical study of finite Larmor radius (FLR) effects in a simplified model of 
$\mathbf{E}\times\mathbf{B}$ transport by drift-waves. The FLR effects are incorporated through the gyro-averaging of the electrostatic potential resulting in an effective drift-wave amplitude proportional to  $\gamma=J_0(\hat{\rho})$ where 
$\hat{\rho}$ is the dimensionless Larmor radius. Based on a weak-turbulence type assumption the drift-wave electrostatic potential is modeled as a superposition of modes that allows to reduce the model to a discrete Hamiltonian dynamical system. This system, known as the gyro-averaged standard map (GSM), generalizes the standard map by introducing the FLR dependence, $\gamma$, on the perturbation amplitude. 

Assuming a  Maxwellian distribution of Larmor radii, we computed the probability density function (pdf), $g$, of the gyro-averaged drift-wave amplitude, $\gamma$. Analytical and Monte-Carlo numerical simulations show that $g(\gamma)$ has singularities at the locations of the extrema of $J_0(\hat{\rho})$. However, depending on the value of the thermal Larmor radius, $\rho_{th}$, these singularities can be exponentially damped in the case of a Maxwellian distribution of Larmor radii. 
Intuitively speaking,  the singularities can thus be classified as ``strong'' (showing clear peaks) or ``weak'' (not showing clear peaks) depending on the role of the Maxwellian exponential damping  factor. 

Results were also presented on the statistical moments of $\gamma$, and it was observed that
 the average of $\gamma$ decays monotonically exponentially fast with $\rho_{th}$, while 
the dispersion of $\gamma$  increases from zero to a maximum value and eventually decays for increasing values of $\rho_{th}$.
An analytical formula for the cumulative distribution function (cdf), $G(\gamma)$, was obtained and validated with numerical simulations.
It was shown that $G$ lacks differentiability due to small scale corners located at the singularities of $g$.  
Our interest in $G(\gamma)$ comes from the fact that it allows to obtain formulas for probabilities 
associated with specific values of the effective perturbation.  

Based on the statistics of $g$ and $G$, analytical results and Monte-Carlo numerical simulations were used to perform a systematic study of the transport properties of the system.  In particular, the probability that a given particle in the plasma could in principle escape (i.e. loss of confinement) was computed as a function of the thermal Larmor radius and the drift wave amplitude. The results show clear evidence that FLR effects suppress transport. That is, for a given drift-wave amplitude, the probability that a particle will remain confined increases with the Larmor radius. This behavior is also observed in the escape rate, $\eta_e$, that increases with the drift-wave amplitude but decreases with $\hat{\rho}_{th}$. 

The numerical results show that the probability of global chaos is an upper bound of the escape rate.
The difference between both measures occurs because
the escape rate in the standard map has not a well-defined transition at the critical parameter that defines the transition to global chaos. 
Many orbits, even in the global chaos regime, 
can remain trapped inside stability islands forever or stick to the boundaries of the stability islands for very long periods of time. Despite the difference, which reduces for high values of the perturbation parameter,  
the escape rate can not be greater than the probability of global chaos. A particle can escape only in a global chaos regime, but not all of the particles moving in global chaos phase spaces can do it. 

The statistics of particle trapping was also studied, and it was shown that the probability that a particle will remain trapped in a drift-wave resonance tends to increase when the Larmor radius increases, verifying once more the role of FLR effects in the suppression of transport.  
We showed that the probability of trapping is a good theoretical estimate for the rate of trapping in the GSM model.
This is due to a well-defined transition in the standard map's rate of trapping near the fixed point's stability boundary.
Thus, in the context of GSM model, this well-defined transition makes the rate of trapping 
more strongly related to the statistics of the effective perturbation.

The analytical results presented in this paper can also be applied to other transport measures, e.g.
diffusion coefficients, often studied in the highly simplified context of the  standard map (i.e., without FLR effects).
Combining known transport properties of the standard map and the results obtained here, we can investigate further transport properties of the GSM model, including non-diffusive transport processes. 
Other possible direction for future studies includes the application of the methodology presented here, that combines 
statistics and nonlinear dynamics, to more sophisticated  gyro-averaged $\mathbf{E}\times\mathbf{B}$ models.

\section{Acknowledgments}
This work was made possible through financial support from  
the S$\tilde{\rm a}$o Paulo Research Foundation (FAPESP, Brazil) under grants No 2011/19296-1,
CNPq (grant  203460/2014-6), FAPESP (grant 2012/10240-6) and DFG (IRTG 1740).
JDF acknowledges Roberto Venegeroles (UFABC, Brazil) for valuable discussions
and the hospitality of the Institute of Physics at the Humboldt University in Berlin, where part of the work was conducted.
DdcN acknowledges support from the Office of Fusion Energy Sciences 
of the US Department of 
Energy at Oak Ridge National Laboratory, managed by UT-Battelle, LLC, 
for the U.S.Department of Energy under contract DE-AC05-00OR22725.


\end{document}